\documentclass[pra,showpacs,a4paper,%
superscriptaddress,floatfix]{revtex4}

\usepackage{graphicx}
\usepackage{amsmath,amssymb}

\begin{document}

\title{Electron attachment to SF$_6$ and lifetimes of SF$_6^-$ negative ions}

\author{L. G. Gerchikov}
\email[Email: ]{lgerchikov@rambler.ru}
\affiliation{St Petersburg State Polytechnic University, 195251,
St Petersburg, Russia}
\author{G. F. Gribakin}
\email[Email: ]{g.gribakin@qub.ac.uk}
\affiliation{Department of Applied Mathematics and Theoretical Physics,
Queen's University, Belfast BT7 1NN, Northern Ireland, United Kingdom}

\date{\today }

\begin{abstract}
We study the process of low-energy electron capture by the SF$_6$ molecule.
Our approach is based on the model of Gauyacq and Herzenberg [J. Phys. B
{\bf 17}, 1155 (1984)] in which the electron motion is coupled to the
fully symmetric vibrational mode through a weakly bound or virtual $s$ state.
By tuning the two free parameters of the model, we achieve an accurate
description of the measured electron attachment cross section and good
agreement with vibrational excitation cross sections of the fully symmetric
mode.
An extension of the model provides a limit on the characteristic time of
intramolecular vibrational relaxation in highly-excited SF$_6^-$. By
evaluating the total vibrational spectrum density of SF$_6^-$, we estimate
the widths of the vibrational Feshbach resonances of the long-lived negative
ion. We also analyse the possible distribution of the widths and its effect
on the lifetime measurements, and investigate nonexponential decay
features in metastable SF$_6^-$.
\end{abstract}

\maketitle


\section{Introduction}\label{sec:intro}

Electron attachment to SF$_6$ is a fascinating problem. In spite of a lot
of attention and a wealth of experimental data \cite{CO01}, some basic
questions, e.g. that of lifetimes of metastable SF$_6^-$, lack definitive
answers. In this paper we show that the electron capture process is
described well by a zero-range-type model \cite{Gauyacq}. We determine the
parameters of the model by comparing the theory with experimental data on
the attachment, total and vibrational excitation cross sections. We then
study the autodetachment widths of SF$_6^-$, and analyze its lifetimes
and nonexponential decay. Here the experimental situation is less clear.
Our calculation yields characteristic lifetimes of about a millisecond,
using possibly the most accurate set of SF$_6^-$ vibrational frequencies
\cite{GB98}. We investigate the nature of nonexponential decay of metastable
anions due to level-to-level fluctuations of the widths and a
distribution of the incident electron and target energies.

Sulfur hexafluoride (SF$_6$) has long been know as an electron scavenger
because of its large low-energy electron attachment cross section. This
feature of SF$_6$ is important for its applications as a gaseous dielectric
and makes for the rich physics of low-energy electron collisions with it.
The energy dependence of the electron capture cross section is well
established experimentally \cite{CO01,CA85,KRH92,HKD01,Hotop,BBM05,Fabrikant}.
Below 10 meV it shows $1/v$ behavior characteristic of $s$-wave inelastic
processes. At higher energies towards 100 meV it approaches
60\% of the unitary limit $\pi /k^2$ for the reaction cross sections,
where $k$ is the incident electron momentum (atomic units are used throughout).
In addition, SF$_6$ also has a large elastic scattering cross section
reaching $\sim 10^3$ a.u. at electron energies of a few meV, which can be
inferred from the measured total cross section \cite{FRS82,Field}.
Note though that experimental data on low-energy elastic
collisions, including differential cross sections \cite{Rohr1}, and inelastic
scattering cross sections, e.g. those of vibrational excitations
\cite{Fabrikant,Rohr,Field1}, are relatively scarce.

Low-energy electron attachment leads primarily to the formation of long-lived
parent negative ions,
\begin{equation}\label{eq:att}
e^-+\mbox{SF}_6\rightarrow \mbox{SF}_6^{-*},
\end{equation}
At electron energies $\varepsilon > 0.2$ eV (and below 3 eV)
dissociative attachment,
\begin{equation}\label{eq:disatt}
e^-+\mbox{SF}_6\rightarrow \mbox{SF}_5^- +\mbox{F},
\end{equation}
becomes dominant. Although this process is usually regarded as a channel
separate from (\ref{eq:att}), some recent evidence suggests that SF$_5^-$
can be formed in the decay of metastable SF$_6^-$ \cite{BRH06,GF07},
\begin{equation}\label{eq:dissoc}
\mbox{SF}_6^{-*}\rightarrow \mbox{SF}_5^- +\mbox{F}.
\end{equation}
However, at low energies the lifetimes of SF$_6^{-*}$ in the absence of
collisions are limited by electron autodetachment,
\begin{equation}\label{eq:detach}
\mbox{SF}_6^{-*}\rightarrow \mbox{SF}_6 + e^-.
\end{equation}
Numerous SF$_6^{-*}$ lifetime measurements
\cite{EG62,CC66,HT71,DA86,AD88,Garrec,Henis,Odom,Dunning,Dunning1}
show considerable variation, depending on the technique used.
Time-of-flight experiments \cite{EG62,CC66,HT71,DA86,AD88,Garrec}
typically yield values of several tens of microseconds, while
ion-cyclotron-resonance methods \cite{Henis,Odom,Dunning,Dunning1} give
values of about a millisecond or larger.

Using the ion-cyclotron-resonance method,
Odom {\em et al.} \cite{Odom} found that the apparent lifetime of
SF$_6^{-*}$ varied as a function of the observation time, and surmised that
SF$_6^-$ were formed in a distribution of states with different lifetimes.
The most recent study that used a Penning ion trap \cite{Dunning1} also
points to the formation of ions with a range of 1--10~ms lifetimes.
A similar conclusion was derived from the time-of-flight measurements
by Delmore and Appelhans \cite{DA86,AD88} at microsecond time scales.
Their analysis of the SF$_6^{-*}$ decay indicated multiple lifetimes or
groups of lifetimes in the interval between 2 to 30~$\mu $s, and showed
that the population of states with
different lifetimes depended on the temperature of SF$_6$ molecules.
On the other hand, Ref. \cite{Garrec} where a free jet expansion was used
to cool down the SF$_6$ molecules, reported a single lifetime of
$19.1\pm 2.7~\mu$s. All these differences are usually attributed to the
differences in the experimental conditions under which SF$_6^{-*}$ are formed,
i.e., the incident electron energy and the internal energy of the target
molecule \cite{CO01}. However, detailed understanding is still lacking.

In contrast to the problem of lifetimes, the process of low-energy electron
attachment to SF$_6$ is described well by a simple zero-range-type
model of Gauyacq and Herzenberg \cite{Gauyacq}. According to this model, the
incoming $s$-wave electron undergoes strong resonant scattering on a
virtual or weakly bound level of the SF$_6$ molecule, with a
near-zero energy. Electron trapping occurs via population of this fully
symmetric state which is strongly coupled to the symmetric stretch
(``breathing'') vibrational mode $\nu _1$. Electron capture initiates the
motion of the fluorine nuclei towards the equilibrium configuration of
the negative ion. This process is accompanied by rapid intramolecular
vibrational redistribution (IVR) of the breathing mode energy among other
vibrational degrees of freedom.
As a result, the probability for the nuclei to return to the equilibrium
configuration of the neutral SF$_6$ becomes small, and long-lived metastable
anions are formed.

In this paper we perform a comprehensive study of electron attachment using
the zero-range model.
The attachment cross section is sensitive to the behavior of the
SF$_6$ and SF$_6^-$ potential energy curves, as a function of the S--F bond
length, near their merging point. Details of this behavior are incorporated
in the model via two parameters, namely the energy of the virtual (or weakly
bound) level and its coupling to the breathing mode.
We develop a new effective matching procedure that
connects the region of electron capture near the merging point,
where the zero-range model can be applied, with the outer region of adiabatic
semiclassical nuclear motion. We also generalize the model to
include the possibility of the nuclear framework to oscillate back to its
initial configuration, in order to study the influence of the rate of IVR
on $e^-+\mbox{SF}_6$ collisions.
Our aim here is to test the model by comparison with experimental data
on the attachment, total and vibrational excitation cross sections, and
thus determine its parameters. Another goal is to compare these parameters
with the results of potential curve calculations.
The SF$_6$ and SF$_6^-$ potential curves have been established quite
well by now \cite{GB98,Tachikawa}, overcoming earlier uncertainties
\cite{Hay82,KBS87,RB93,KGS96,KDG97}.

A somewhat different theoretical approach was taken recently
by Fabrikant {\em et al.} \cite{Fabrikant}. It starts from the $R$-matrix
formalism, and goes beyond the model of Gauyacq and Herzenberg by
including the long-range polarization potential $-\alpha /2r^4$ and dipole
coupling between the $s$ and $p$ waves. Contributions of higher electron
partial waves are also included, to describe the elastic and vibrational
excitation cross sections more accurately. In this theory the two parameters
of the $s$-wave coupling to the $\nu _1$ mode are allowed to be complex.
Their values are found by fitting the experimental attachment and total
cross sections. However, the use of complex parameters makes them
phenomenological, and the physical connection with the molecular potential
curves is lost.

The second part of our paper concerns the evaluation of lifetimes
of SF$_6^{-*}$ due to electron autodetachment.
The first estimates of the metastable anion lifetimes and their relation to
the attachment cross section, vibrational spectrum density of SF$_6^-$
and electron affinity of SF$_6$ were made in Ref. \cite{CC66}.
The rate constants of the processes (\ref{eq:dissoc}) and (\ref{eq:detach}),
were later studied \cite{Klots,Klots1,Lifshitz,Weston} using the
quasi-equilibrium (or RRKM) theory \cite{Forst}.
The approach of Refs. \cite{CC66,Klots,Klots1} is based on the principle
of detailed balance, and requires the knowledge of the attachment cross section
and the anion vibrational spectrum density. The standard RRKM approach is
similar, assuming in addition that the transition probability is unity,
i.e., that the attachment cross section is equal to its unitary limit.
On the other hand, the RRKM requires the knowledge of the density of so-called
transition states. The autodetachment lifetime depends strongly on such
parameters as the electron affinity $E_a$ and vibrational spectrum density of
SF$_6^-$, which were not known well. Using different sets of data, lifetimes
from microseconds to milliseconds were obtained in a wide range of
incident electron energies. It should be noted that these methods 
yield the detachment rates (inverse lifetimes) {\em averaged} over
the ensemble of metastable anion states. The distribution of lifetimes within
such ensemble that may cause a variation in the observed lifetimes, has
not been studied.

In the present paper we analyze the dependence of the autodetachment rate
on the incident electron energy in the interval from zero to 300 meV for
different target temperatures. Our calculations are based on the accurate
values of the attachment cross section, and take into account the
distribution of the target SF$_6$ molecules over the rotational and
vibrational states, as well as the rotational and vibrational degrees of
freedom of the SF$_6^-$ anion. We also analyse fluctuations of the decay
rate over the ensemble of metastable SF$_6^-$, to see if they can explain
observations of nonexponential decay.


\section{Electron scattering and capture}

\subsection{Attachment model}\label{subsec:att}

Following the approach of Gauyacq and Herzenberg \cite{Gauyacq} we
employ the zero-range potential (ZRP) model to describe the electron 
interaction with SF$_6$. ZRP theory is a well-known tool suitable for
problems of low-energy electron scattering and negative ions
\cite{Demkov,DO88}, especially in those cases when the cross section is
enhanced by the existence of a shallow bound state or low-lying virtual level
with zero angular momentum (i.e., in the $s$ wave). Application of the ZRP
method to the $e^-+\mbox{SF}_6$ system provides the wavefunction for the
continuous spectrum electron and SF$_6$ in the vicinity of its equilibrium
configuration.
On the other hand, the wavefunction of SF$_6^-$ formed as a result
of electron attachment can be described in the Born-Oppenheimer approximation,
treating the nuclear motion semiclassically. Matching of these wavefunctions
yields the electron attachment cross section together with those of elastic
scattering and vibrational excitation.

The electronic state responsible for the capture process is a fully symmetric
weakly bound state of $e^-+\mbox{SF}_6$. Outside the molecule the
electron wavefunction $\psi _{0}$ is spherically symmetric, and is given by
\begin{equation}\label{WF0}
\psi _{0}=\sqrt{\frac{\kappa }{2\pi }}\frac{e^{-\kappa r}}{r},
\end{equation}
where $r$ is the electron coordinate and $\kappa $ is related to the
bound state energy $\varepsilon _0=-\kappa ^2/2$. 
The energy $\varepsilon _0$ depends on all nuclear coordinates. However,
due to its symmetry, this electron state is most strongly coupled to
the breathing (symmetric stretch) vibrational mode of SF$_6$. This allows
one to consider $\varepsilon _{0}$ and $\kappa $ as functions of the normal
coordinate $q$ of the breathing mode, $q=R-R_0$, where $R$ is the S--F
bond length, and $R_0$ is its value at the equilibrium of SF$_6$.
Near the equilibrium, $\kappa (q)$ can be expanded in a power series.
Keeping the first two terms \cite{Gauyacq},
\begin{equation}\label{kapa(q)}
\kappa (q)=\kappa _0+\kappa _1q,
\end{equation}
where $\kappa _1>0$, since the binding increases with the increase in $R$
($R=3.25$ a.u. at the equilibrium of SF$_6^-$, while
$R_0=2.96$ a.u. \cite{GB98}).

\begin{figure}[tbp]
\includegraphics*[width=10cm]{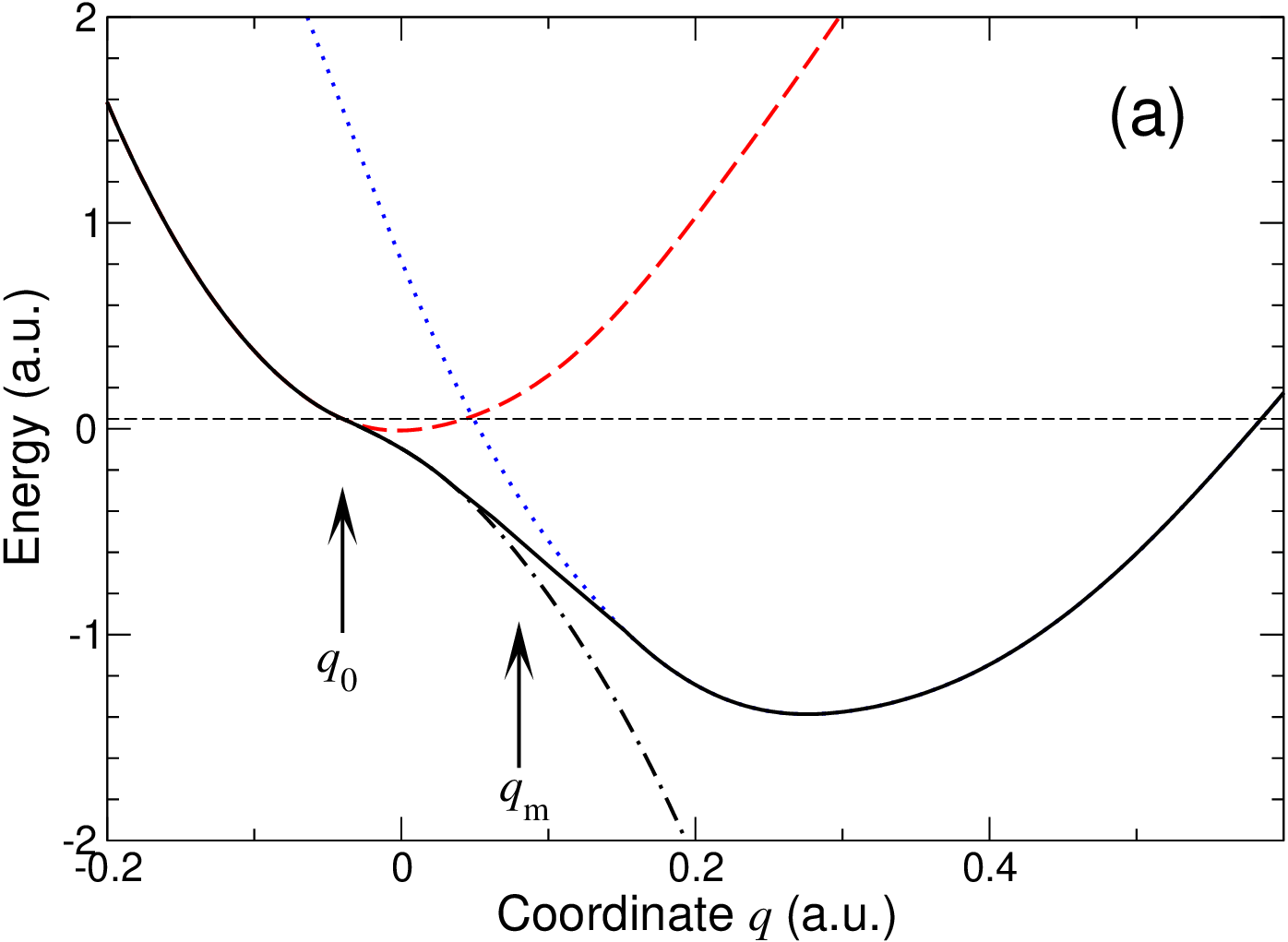}

\includegraphics*[width=10cm]{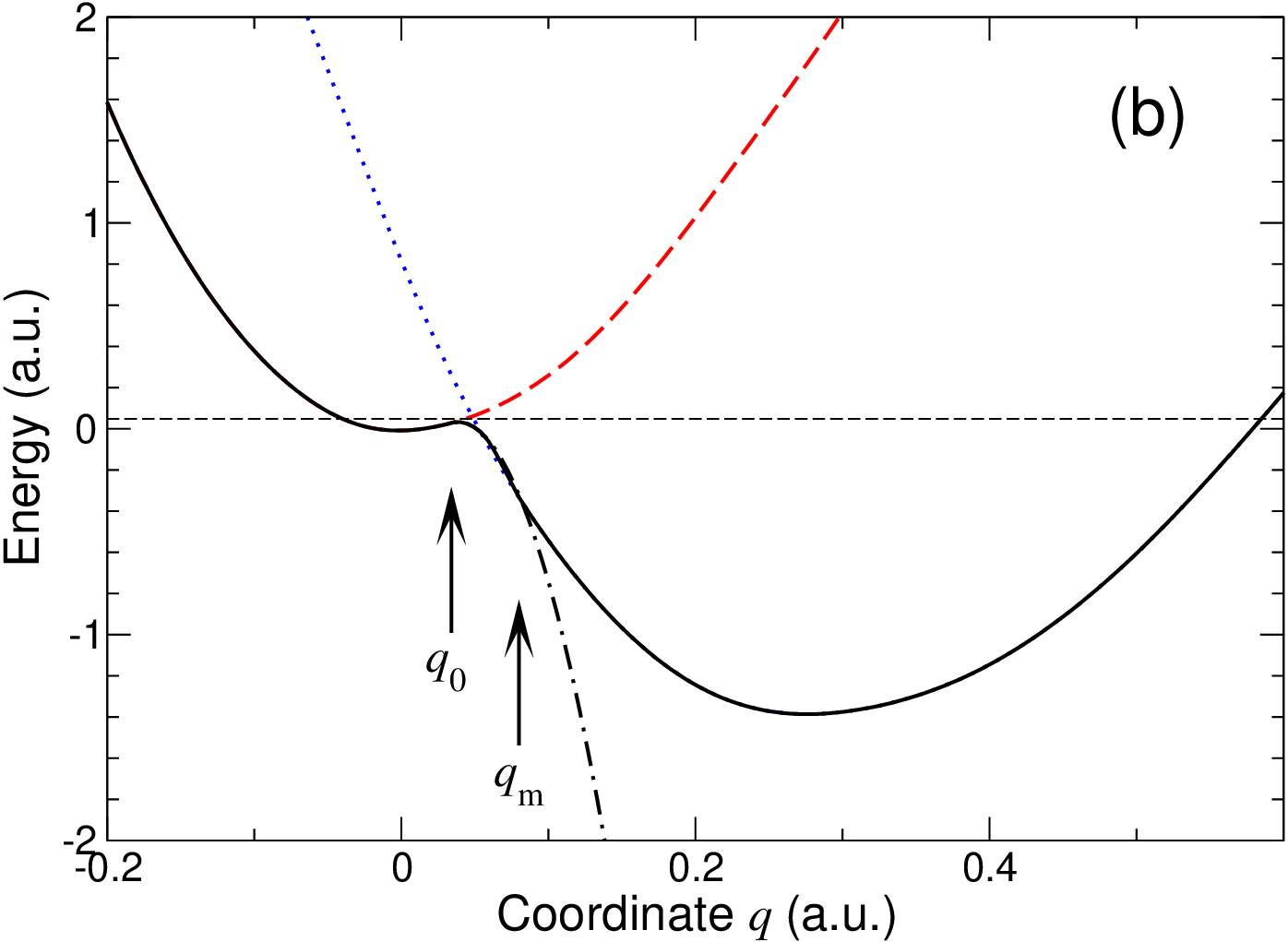}
\caption{SF$_6$ and SF$_6^-$ potential curves for the symmetric stretch
coordinate. Dashed and dotted curves are the SF$_6$ and SF$_6^-$
energies, respectively, calculated using the 2nd order M\o ller-Plesset
perturbation theory \cite{Tachikawa}. Dot-dashed curves show the SF$_6^-$
energy in the quadratic approximation for $q_0=-0.04$, $\kappa _1=2.0$ (a)
and $q_0=0.034$, $\kappa _1= 4.1$ (b). Solid curves interpolate between
the quadratic and numerical SF$_6^-$ potential curves and represent the
lowest adiabatic energy of the $e^-+\mbox{SF}_6$ system. Horizontal dashed
lines show the lowest total energy of $e^-+\mbox{SF}_6$ collision,
$\omega /2$.}
\label{fig:pec}
\end{figure}

The negative ion bound state (\ref{WF0}) exists when $\kappa >0$, i.e., for
$q>q_0\equiv -\kappa _0/\kappa _1$, while $\kappa <0$ corresponds to a virtual
state \cite{Landau}. The absolute value of $\kappa _0$ is expected to be small,
$|\kappa _0| \ll 1$~a.u., because for the electron capture to be effective,
the scattering length $\kappa ^{-1}$ should be large in the Frank-Condon
region. The sign of $\kappa _0$ indicates whether the negative ion state is
real or virtual at the equilibrium of the neutral. The two cases are
illustrated in Fig. \ref{fig:pec}, which shows the potential curves of
SF$_6$ and SF$_6^-$ for the symmetric stretch coordinate.

To the right of the merging point, $q_0$, indicated in Fig. \ref{fig:pec} by
vertical arrows, the negative ion energy $U_{0}(q)$ is given by the
sum of that of the neutral SF$_6$ and the bound state energy
$\varepsilon _{0}(q)=-\kappa ^{2}(q)/2$. The SF$_6$ potential curve shown 
was calculated in Ref. \cite{Tachikawa}. Near the equilibrium
it can be approximated by $\frac 12 M\omega ^2q^2$, where $\omega $ and
$M$ are the frequency and mass of the breathing mode. Using
Eq. (\ref{kapa(q)}) one obtains a quadratic approximation for the SF$_6^-$
potential curve near the merging point,
\begin{equation}\label{eq:U_0}
U_0(q)\simeq \frac {M\omega ^2q^2}{2}-\frac{(\kappa _0+\kappa _1q)^2}{2}.
\end{equation}
For $q$ to the far right of the merging point, $U_{0}(q)$ should approach the
anion potential energy curve, e.g., that calculated in Ref. \cite{Tachikawa}.

In this model the electron scattering length $\kappa ^{-1}$ depends
on the breathing vibrational coordinate $q$. This means that the
electron interacts only with this particular mode, and no other vibrations
are excited in the process of electron scattering or capture. Hence,
all nuclear coordinates except $q$ can be omitted, and the the total
wavefunction of the system can be written as
\begin{equation}\label{eq:WF1}
\Psi ({\bf r},q)=e^{i{\bf k}\cdot {\bf r}}\chi _{n_{0}}(q)+
\sum _n \frac{f_n}{r}e^{ik_{n}r}\chi _{n}(q).
\end{equation}
The wavefunction (\ref{eq:WF1}) is an expansion over the SF$_6$ breathing mode
vibrational states $\chi _{n}(q)$, taken in the harmonic approximation, with
energies $E_{n}=\omega (n+\frac{1}{2})$. The first term in Eq.(\ref{eq:WF1})
describes the electron with momentum ${\bf k}\equiv {\bf k}_{n_0}$ incident
on the target in the initial state $n_{0}$. The sum over $n$ accounts
for elastic scattering ($n=n_0$) and vibrationally inelastic processes
($n\neq n_0$). Energy conservation,
$\frac 12 k^{2}+E_{n_{0}}=\frac 12 k_{n}^{2}+E_{n}$, determines the
corresponding electron momenta, $k_{n}=\sqrt{k^{2}-2\omega (n-n_{0}) }$.

Note that the sum in Eq. (\ref{eq:WF1}) includes both open (real $k_{n}$)
and closed ($k_{n}=i|k_{n}|$) channels. In the former,
the electron escapes but the nuclear motion is finite. Closed channels
involve the electron localised near the origin, with the
vibrational motion in progressively higher $n$ states.
Its contribution describes electron attachment, with the nuclei sliding down
the negative ion potential curve towards greater $q$. It takes the form of the
SF$_6^-$ anion adiabatic wavefunction $\Psi _{\rm att}({\bf r},q)$ (see below).
Of course, the true anion potential curve (Fig. \ref{fig:pec}) does not allow
for the infinite nuclear motion, as the nuclear framework swings back to the
neutral equilibrium after one vibrational period. However, in a molecule with
many vibrational degrees of freedom the energy may dissipate
into other modes, providing for electron capture on much longer time scales.

The electronic parts of the wavefunction (\ref{eq:WF1}) are plane or
spherical waves, which ensures its correct asymptotic form. They are
solutions of the free-particle Schr\"odinger equation.
This is in accordance with the ZRP method, in which the potential affects
the wavefunction through the boundary condition at the origin,
\begin{equation}\label{eq:zero}
\left. \frac{1}{r\Psi }\frac{\partial (r\Psi )}{\partial r}
\right|_{r\rightarrow 0}=-\kappa (q).
\end{equation}
Because of the $r\rightarrow 0$ limit, the ZRP affects only
the electron $s$ wave.

Applying Eq. (\ref{eq:zero}) with $\kappa (q)$ from (\ref{kapa(q)}) to
the wavefunction (\ref{eq:WF1}) and projecting the resulting equation onto
each of the nuclear vibrational states $\chi _n$, one obtains a set of linear
equations for the amplitudes $f_n$ ($n=0,\,1\,,\dots $) \cite{Gauyacq}:
\begin{equation}\label{eq:f_n}
(ik_{n}+\kappa _0)f_n+\frac{\kappa _1}{\sqrt{2M\omega }}\left(
\sqrt{n}f_{n-1}+\sqrt{n+1}f_{n+1}\right) =-\delta _{nn_{0}}.
\end{equation}
The general solution of this 2nd-order recurrence relation is a linear
combination of two independent solutions with arbitrary coefficients.
One ``boundary condition'' is provided by Eq. (\ref{eq:f_n}) with
$n=0$. The other boundary condition is set at large $n$. It is related to
the asymptotic behaviour of $\Psi _{\rm att}({\bf r},q)$ at large $q$.
Analysis presented in Appendix \ref{app:match} shows that there is a
simple relation between the nuclear coordinate $q$ and the quantum number
$n$ of the terms in the expansion, Eq. (\ref{eq:WF1}), which contribute
significantly to the wavefunction at this $q$,
\begin{equation}\label{Nq}
n\approx n_0+[k^{2}/2-\varepsilon _{0}(q)]/\omega .  
\end{equation}
Hence, the wavefunction $\Psi ({\bf r},q)$ from Eq. (\ref{eq:WF1}) must
be extended to the region where the incident electron is bound, and
matched at some point $q=q_{\rm m}$ with the negative ion wavefunction
$\Psi _{\rm att}({\bf r},q)$. The choice of $q_{\rm m}$ should not
affect the capture cross section. Physically, it is restricted
to the range of validity of expansion (\ref{kapa(q)}), so $q_{\rm m}$
should not be too large, but sufficient to neglect nonadiabatic effects.

In this region one can write $\Psi _{\rm att}({\bf r} ,q) $ in the
Born-Oppenheimer approximation,
\begin{equation}\label{NI}
\Psi _{\rm att}({\bf r},q)=\psi _{0}({\bf r},q)\chi (q),
\end{equation}
where $\psi _{0}({\bf r},q)$ is the bound electron wavefunction (\ref{WF0}),
which depends on the nuclear coordinate via $\kappa =\kappa (q)$, Eq.
(\ref{kapa(q)}), and $\chi (q)$ is the wavefunction of the nuclear motion
of the anion. It is in general a superposition of the outgoing and
incoming (reflected) waves,
\begin{equation}\label{eq:chi}
\chi (q)=A\chi ^{(+)}(q)+B\chi ^{(-)}(q),
\end{equation}
where $\chi ^{(\pm )}$ can be written explicitly in the semiclassical (WKB)
approximation \cite{Landau}, as $\chi ^{(\pm )}=v^{-1/2}e^{\pm i\int pdq}$,
$p$ and $v$ being the classical nuclear motion momentum and velocity.

The amplitude of reflection, $R=B/A$, depends on the behavior of the negative
ion term $U_{0}(q)$ far from the merging point and also on the coupling
of the anion breathing mode to the other vibrational modes (i.e., on the
effectiveness of IVR). In the absence
of such coupling one has $|R|=1$ and no capture takes place. In treating
electron attachment to SF$_6$, the usual assumption is that this
coupling is strong, i.e. $|R|\ll 1$, so that only the outgoing wave is retained
in Eq.~(\ref{eq:chi}) \cite{Gauyacq}.

In order to find the solutions of Eq. (\ref{eq:f_n}) corresponding to
$\chi ^{(\pm )}(q)$, note that for a sufficiently large $n$ one has:
$ik_n+\kappa _0\simeq -\sqrt{2\omega n}$, and Eq. (\ref{eq:f_n}) turns
into a recurrence relation with constant coefficients. Hence, asymptotically,
the two independent solutions of Eq. (\ref{eq:f_n}) behave as
$f_n\propto \xi ^n$,
where $\xi $ is either of the two complex-conjugate roots of the corresponding
characteristic equation. It turns out (see Appendix \ref{app:match}) that
one of these solutions corresponds to $\chi ^{(+)}$, while the other
to $\chi ^{(-)}$.

In practice, Eq. (\ref{eq:f_n}) approaches the constant-coefficient limit
slowly, as the coefficients vary slowly with $n$ \cite{Braun}. However, this
slowness does allow us to truncate the set (\ref{eq:f_n}) at some large $n=N$
by using $f_{N+1}=\xi f_N$, where $\xi $ satisfies the quadratic equation,
\begin{equation}\label{eq:char}
\kappa _1 \sqrt{\frac{N+1}{2M\omega }}\xi ^2-(|k_N|-\kappa _0)\xi +\kappa _1
\sqrt{\frac{N}{2M\omega }}=0,
\end{equation}
and where $k_N=i\sqrt{2\omega (N-n_0)-k^2}= i|k_N|$ is
imaginary. The two roots of Eq. (\ref{eq:char}) are
\begin{equation}\label{eq:b}
\xi _{\pm }=\frac{\sqrt{M\omega }(|k_N|-\kappa _0)}{\sqrt{2(N+1)}\kappa _1}
\pm i\left[\sqrt{\frac{N}{N+1}}-\frac{M\omega  (|k_N|-\kappa _0 )^2}
{2(N+1)\kappa _1^2}\right]^{1/2}.
\end{equation}
The choice of $N$ is related to the matching point $q_{\rm m}$ by means of
Eq. (\ref{Nq}). The expression in square brackets is real because the
anion potential (\ref{eq:U_0}) must decrease at large $q$, which requires
$\kappa _1^2>M\omega ^2$. Note also that
$|\xi _{\pm }|^2=\sqrt{N/(N+1)}\simeq 1$ for large $N$, which means that
$\xi _{\pm }\simeq e^{\pm i\beta }$ is just a phase factor. 

As follows from Appendix \ref{app:match}, solving the set of $N+1$ equations
(\ref{eq:f_n}) ($n=0,\,1,\dots ,\, N$) with the additional condition
$f_{N+1}=\xi _+ f_N$, generates a set of amplitudes (we denote them
$f_n^{(+)}$) that corresponds to the $A\chi ^{(+)}$ part of the nuclear
wavefunction (\ref{eq:chi}). On the other hand, using $f_{N+1}=\xi _- f_N$,
one obtains a set of amplitudes, $f_n^{(-)}$, which corresponds to
$B\chi ^{(-)}$. There is a simple linear relation between the amplitudes
$f_n^{(+)}$ and $f_n^{(-)}$, and $R$ at large $n$:
$f_n^{(-)}/f_n^{(+)}\equiv R_f= e^{-i\delta _n}R$, where $\delta _n$
is a phase, see Eq. (\ref{eq:R_f}).

In the general case, $R\neq 0$, the solution $f_n$ of the recurrence relations
(\ref{eq:f_n}) that matches $\Psi _{\rm att}({\bf r},q)$,
Eqs. (\ref{NI}), (\ref{eq:chi}), has the form 
\begin{equation}
f_n=C\left( f_n^{(+)}+f_n^{(-)}\right) ,  \label{fn}
\end{equation}
where $C$ is a normalization constant. Therefore, we can set
$f_{N+1}=C(1+R_f)$ and $f_N=C(\xi _+^{-1}+R_f\xi _-^{-1})$.
The amplitudes $f_n$ for $n=N-1,\dots ,\,n_0$ are then found successively
from Eq. (\ref{eq:f_n}). The same is done for $f_n$ with $n=0,\dots ,n_0$,
starting from an arbitrary $f_0$, e.g., $f_0=1$. The two solutions are
then matched at $n=n_0$, and the overall normalization is determined by
substitution in the inhomogeneous equation (\ref{eq:f_n}) with $n=n_0$.

In practice, Eq. (\ref{eq:f_n}) can be truncated at relatively low $n=N$,
e.g., $N\sim 10$ can be used for electron energies below 350 meV without
significant error. By means of Eq. (\ref{Nq}), this allows us to 
keep a physically meaningful value of the matching coordinate,
$q_{\rm m}=0.08$, see Fig. \ref{fig:pec}.

The elastic ($n=n_0$) and vibrationally inelastic ($n\neq n_0$)
cross sections are given by
\begin{equation}\label{sigma_n}
\sigma _{n}=\frac{4\pi }{k}k_{n}\left| f_n\right| ^{2},
\end{equation}
where $k_{n}$ is real for the open channels. The total cross section
is given by the optical theorem \cite{Landau}, as 
\begin{equation}\label{opttot}
\sigma _{t}=\frac{4\pi }{k}\mbox{Im}\,f_{n_{0}} ,
\end{equation}
and the attachment cross section is obtained as
\begin{equation}\label{optDA}
\sigma _{\rm att}=\frac{4\pi }{k}\Bigl( \mbox{Im}\,f_{n_{0}}
-\sum _n |f_n|^2\mbox{Re}\,k_{n} \Bigr) .
\end{equation}
It can also be found directly from the asymptotic behaviour of
$f_n$ for closed channels, Eq.~(\ref{eq:DA}).

Note that in comparison with Ref. \cite{Gauyacq}, our method allows one to
account for nonzero reflection, i.e., for incomplete vibrational relaxation,
and provides the cross sections of all processes (not just attachment).

\subsection{Numerical results}\label{subsec:numres}

The model described in Sec. \ref{subsec:att} contains five parameters.
The frequency of the SF$_6$ breathing mode is well established experimentally,
$\omega =96~{\rm meV}=3.5\times 10^{-3}$~a.u. \cite{CO01}, and the
corresponding mass is $M=6m_{\rm F}=2.1\times 10^5$~a.u. The other parameters
of the model, namely $\kappa _0$ and $\kappa _1$, which characterise the
anion potential curve, and the reflection amplitude $R$, are not known
{\em a priori}.

To determine their values and to verify the model itself, we have performed
numerical calculations of the cross sections in a wide
range of parameters, and compared the results with the experimental data on
attachment \cite{KRH92,CO01,BBM05}, total scattering \cite{Field},
and vibrational excitations \cite{Fabrikant}. Our calculations have been done
for the target molecules in the ground vibrational state of the breathing
mode (i.e., $n_0=0$), since at room temperatures the fraction of excited
states of this mode is small. 

The ZRP model is expected to work best at low electron energies. Here
the attachment data is the most accurate of all the measured SF$_6$ cross
sections, and we use it as our main guide in the search for the optimal
parameters. To characterise the discrepancy between the
theory and experiment we calculate the ``mean-squared relative error'',
$\eta =\sum_{i=1}^{36}[\Delta \sigma _{\rm att}(\varepsilon _i)/
\sigma _{\rm att}(\varepsilon _i)]^{2}/36$, where $\Delta \sigma _{\rm att}$
is the difference between the theoretical and experimental cross sections,
using 36 energies $\varepsilon _i$ between 0.1 and 160 meV, as given in the
tables of recommended cross sections in Ref. \cite{CO01}. Let us first examine
the results obtained for $R=0$ (rapid IVR), and then look at the effect of
the reflected wave, $R\neq 0$ (incomplete IVR).

\subsubsection{Rapid IVR ($R=0$)}\label{sssec:Req0}

\begin{figure}[tbp]
\includegraphics*[width=9cm,angle=-90]{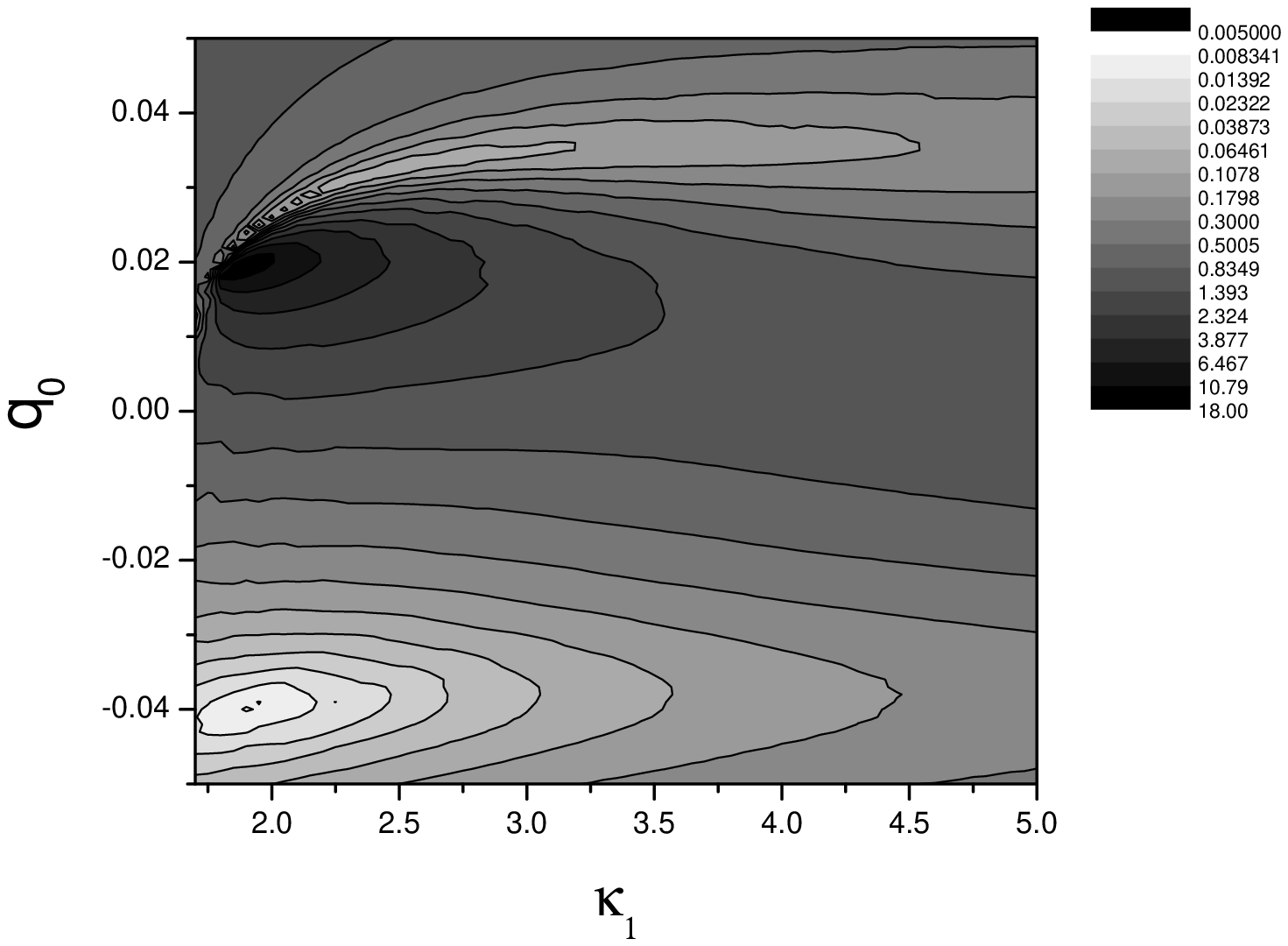}
\caption{Density plot of the mean-squared relative error $\eta $ of the
theory fit of the experimental cross section of electron attachment
to SF$_6$ over the 0.1--160 meV energy range. Lighter areas mean better
fits.}
\label{fig:opt}
\end{figure}

The values of $\eta $ for $R=0$ are shown on a density plot in
Fig.~\ref{fig:opt}, as a function of $\kappa _1$ and
$q_0=-\kappa _0/\kappa _1$. The range of parameters is limited by
$\kappa _1>\sqrt{M}\omega =1.6$~a.u., and the use of $q_0$ instead of
$\kappa _0$ emphasizes the sensitivity of the attachment cross section to
the position of the merging point of the neutral and anion potential curves.

Figure \ref{fig:opt} shows that the absolute minimum of $\eta $ is achieved
for $q_{0}\approx -0.04$ and $\kappa _1 \approx 2$. A negative value of
$q_{0}$ means that the electron does form a weakly bound SF$_6^-$ state at
the equilibrium of the neutral, $q=0$. Figure \ref{fig:pec}(a) shows 
the corresponding anion potential curve, $U_{0}(q)$ from Eq. (\ref{eq:U_0}),
by the dot-dashed curve. For comparison we also show the SF$_6^-$ potential
curve calculated in \cite{Tachikawa}. This
potential curve does not reproduce $U_{0}(q)$ near the merging point
\cite{comment1} but should be reasonably accurate at larger $q$.
Figure \ref{fig:pec} also shows that the analytical and numerical potential
curves can be matched. The position of the matching point $q_{\rm m}=0.08$
used in our calculations is indicated by an arrow. Note though, that for
$R=0$ the calculation does not require any information about the anion
potential curve far from the merging point, and the choice of the matching
point (or the truncation number $N$) is not critical.

\begin{figure}[tbp]
\includegraphics*[width=11cm]{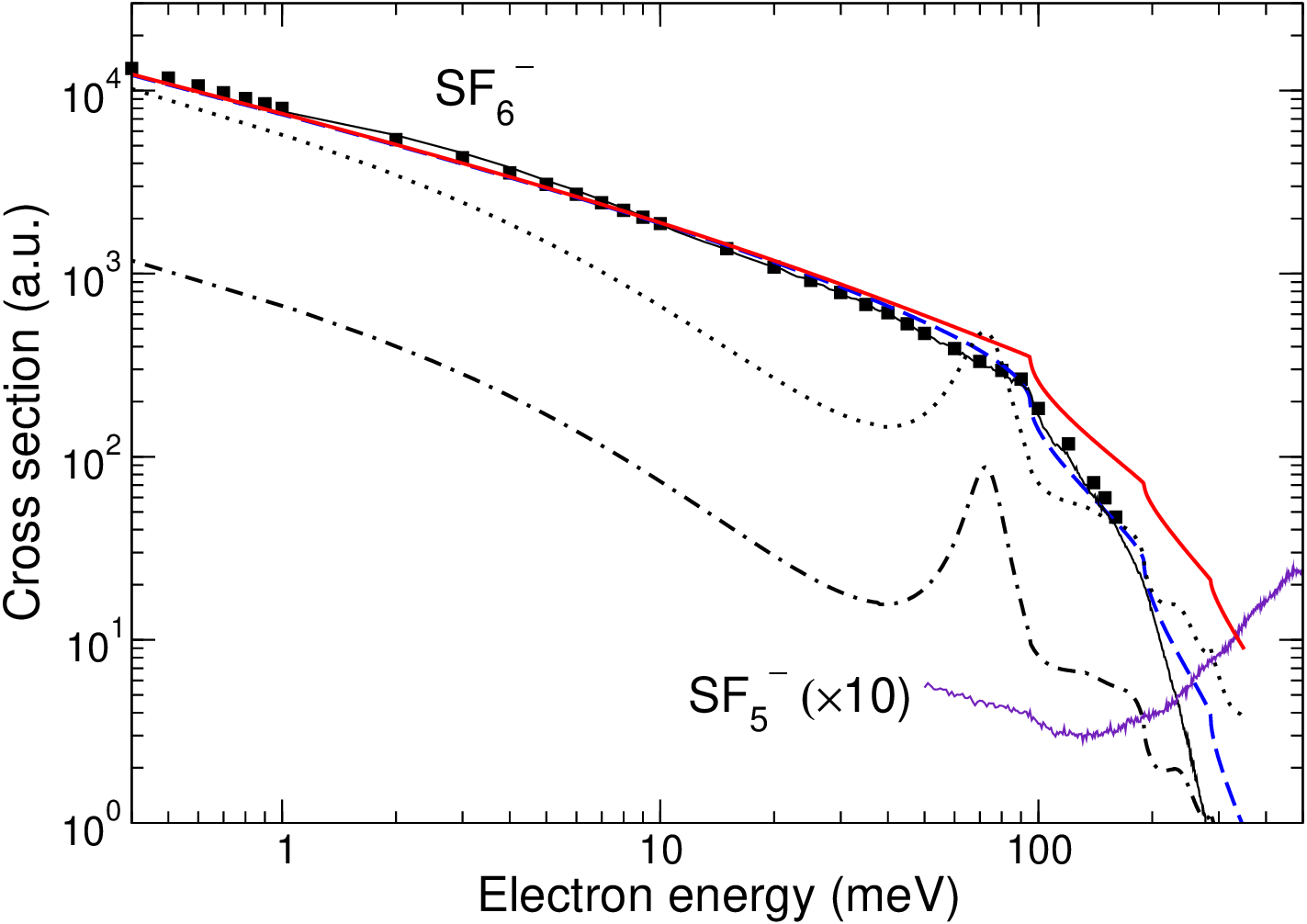}
\caption{Electron attachment cross sections for SF$_6$. Solid squares are
the experimental data for metastable SF$_6^-$ obtained by Hotop {\em et al.}
as given in Ref. \cite{CO01}; thin solids curves show the data for SF$_6^-$
and SF$_5^-$ obtained at the SF$_6$ nozzle temperature of 300 K \cite{BBM05}.
Calculations: $q_{0}=-0.04$, $\kappa _1=2.0$, $R=0$ (dashed curve);
$q_{0}=0.034$, $\kappa _1=4.1$, $R=0$ (thick solid curve),
$R\neq 0$, $\gamma =0.1\omega _a$ (dotted curve), and $\gamma =0.01\omega _a$
(dot-dashed curve).}
\label{fig:att}
\end{figure}

\begin{figure}[tbp]
\includegraphics*[width=11cm]{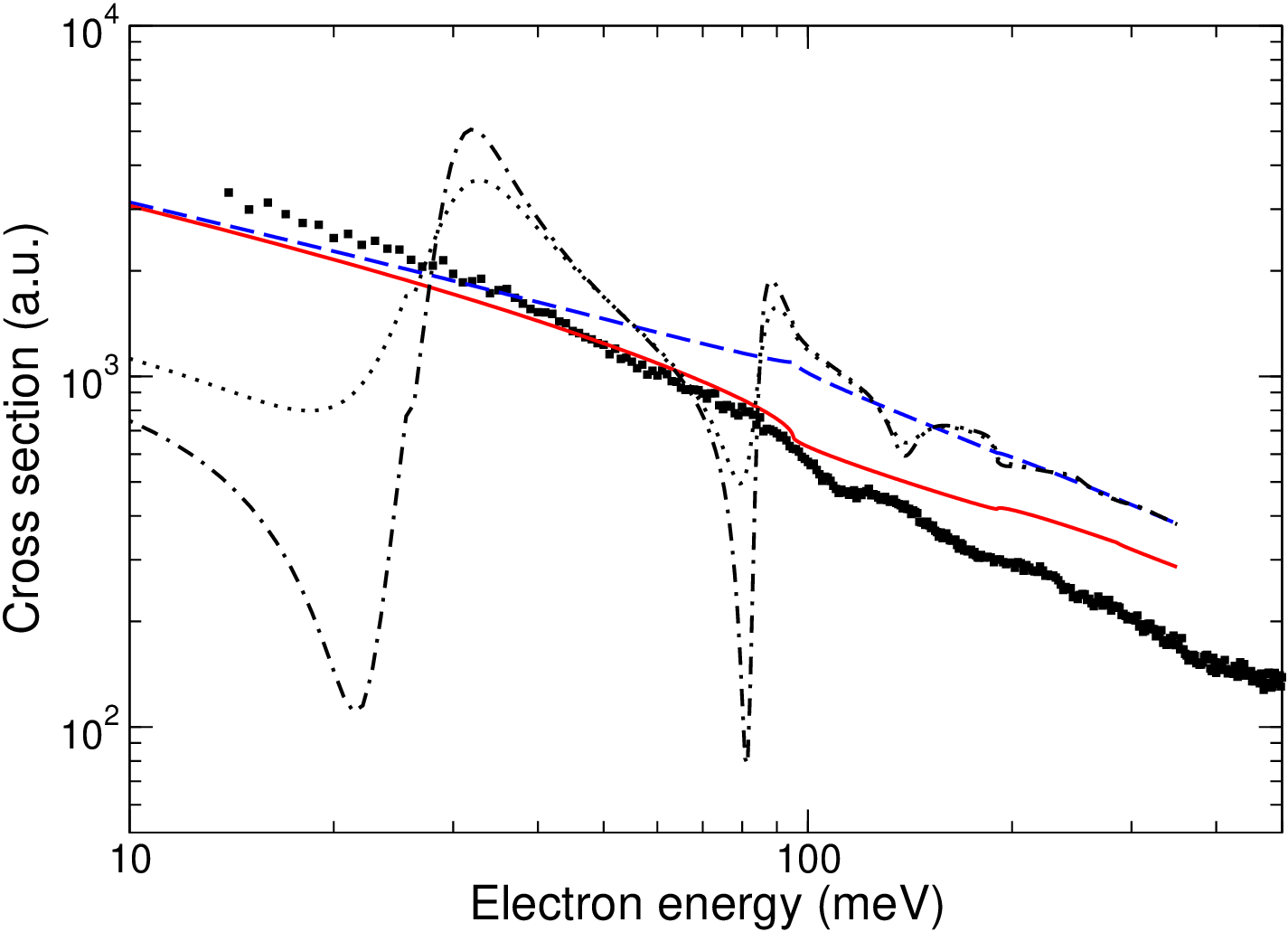}
\caption{Total cross section for electron scattering from SF$_6$. Experimental
data: solid squares, Ref. \cite{Field}.
Calculations: $q_{0}=-0.04$, $\kappa _1=2.0$, $R=0$ (dashed curve);
$q_{0}=0.034$, $\kappa _1=4.1$, $R=0$ (thick solid curve),
$R\neq 0$, $\gamma =0.1\omega _a$ (dotted curve), and $\gamma =0.01\omega _a$
(dot-dashed curve).}
\label{fig:tot}
\end{figure}

The attachment cross section calculated for $q_{0}=-0.04$, $\kappa _1=2.0$
and $R=0$ (dashed curve in Fig. \ref{fig:att}) reproduces both the smooth
decrease of the measured cross section below the $\nu_1$ vibrational threshold,
$\varepsilon <\omega $ \cite{comment2}, and its rapid drop for
$\varepsilon >\omega $. However, the total cross section calculated with these
parameters is noticeably higher than experiment for $\varepsilon
\gtrsim \omega $ (see Fig. \ref{fig:tot}).

The other possibility suggested by Fig. \ref{fig:opt} is that $q_0>0$,
where $\eta $ has a second local minimum in the form of a narrow ``valley''.
It corresponds to a virtual electron level at the equilibrium of SF$_6$.
Here the quality of the fit is not very sensitive to the precise value of
$\kappa _1$. Choosing $q_{0}=0.034$, $\kappa _1=4.1$ and $R=0$ (thick solid
curves in Figs. \ref{fig:att} and \ref{fig:tot}) we obtain a much better
description of the total cross section, while the fit of the attachment
data is only slightly worse. In Fig. \ref{fig:pec}(b) we show the
corresponding behavior of the SF$_6^-$ potential curve near the equilibrium.
In this case the quadratic curve matches the SF$_6^-$ curve from
Ref. \cite{Tachikawa} even closer at $q\approx q_{\rm m}=0.08$. Note though,
that due to the proximity of $q_{\rm m}$ to $q_0$, the adiabatic approximation
for the nuclear motion is not as accurate here as in the $q_0<0$ case (the
required truncation $N$ being lower). This means that the actual shape of
the SF$_6^-$ potential curve in the vicinity of $q_{\rm m}$ may have a small
effect on the cross section.

Generally, the ZRP model is expected to provide
a more accurate description of the attachment cross section than the total
cross section. Due to the symmetry of the negative ion state and the role
played by the symmetric stretch mode, the low-energy attachment model can be
restricted to the electron $s$ wave.
However, contributions of higher partial waves to the total cross section
(in particular, due to excitation of the strong infrared-active $\nu _3$ mode),
may become sizeable even at low energies. This is indicated by the
observed anisotropy of the total cross section \cite{Field} (see also
Ref. \cite{Fabrikant}). Since the present ZRP model does not take into account
any of these contributions, one could expect that the cross section
(\ref{opttot}) would be a lower bound for the true total cross section.
Hence, we believe that the parameters $q_{0}=0.034$ and $\kappa _1=4.1$ are
more realistic, in spite of a less accurate fit of the attachment cross
section than that for $q_{0}=-0.04$, $\kappa _1=2.0$ .

\begin{figure}[tbp]
\includegraphics*[width=11cm]{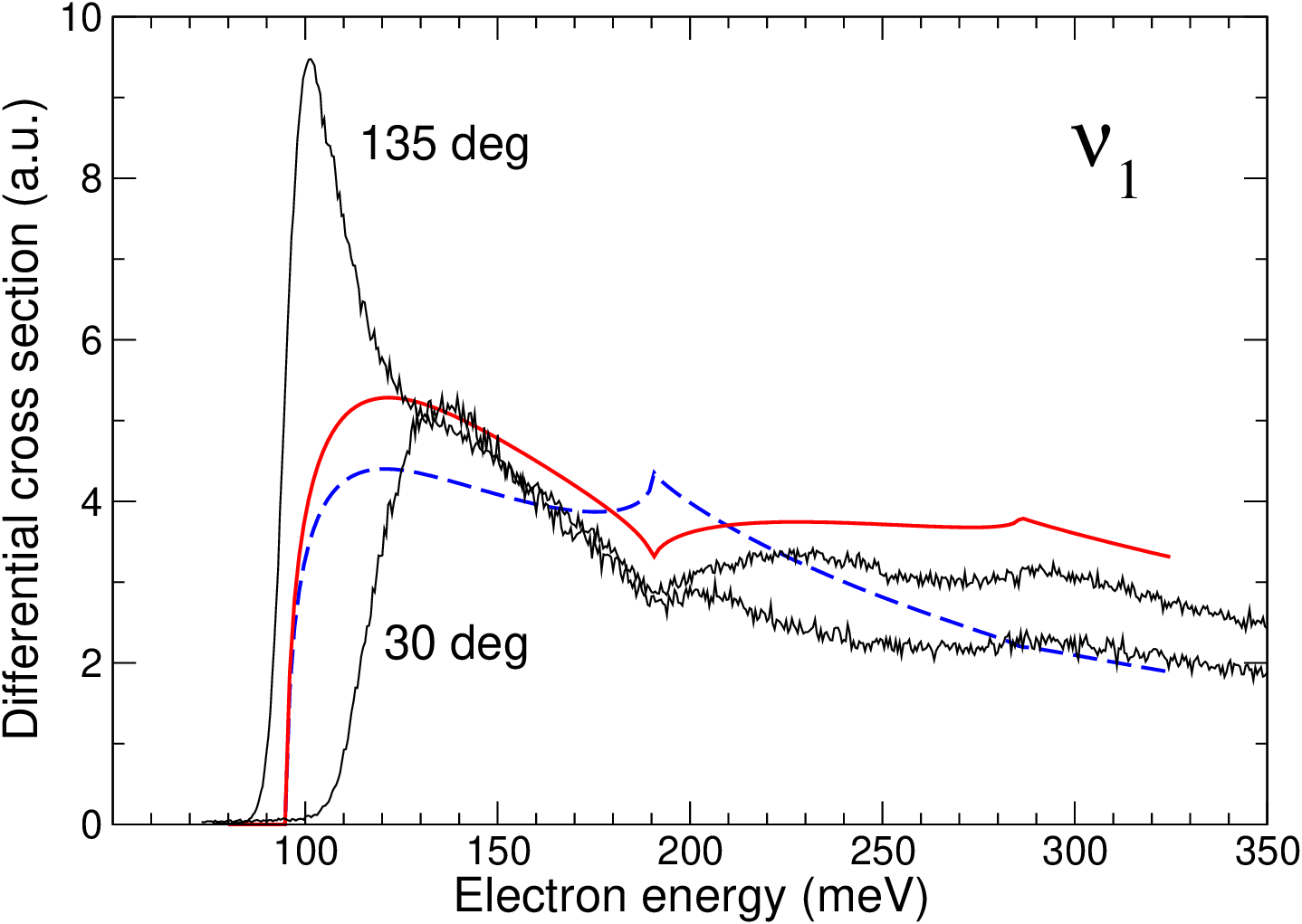}
\includegraphics*[width=11cm]{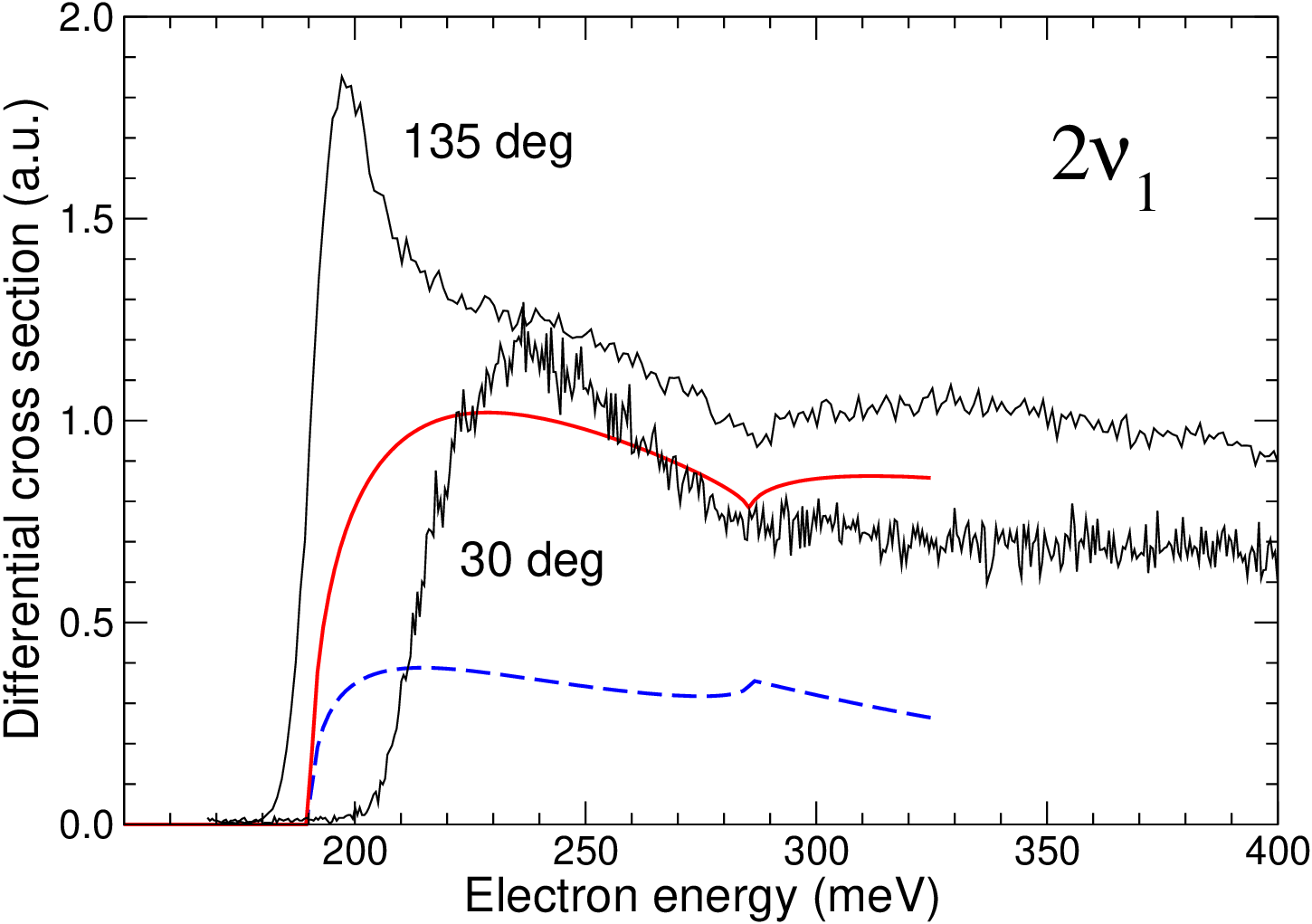}
\caption{Differential vibrational excitation cross sections of the symmetric
stretch mode of SF$_6$ (top, $\nu _1$; bottom, $2\nu_1$). Thin solid lines are
the experimental results of Ref. \cite{Fabrikant} obtained at 30$^\circ$
and  135$^\circ$. Dashed curves are the cross sections calculated
for $q_{0}=-0.04$ and $\kappa _1=2.0$, and thick solid curves are those for
$q_{0}=0.034$ and $\kappa _1=4.1$ (both for $R=0$).}
\label{fig:vib}
\end{figure}

As a final check of the model for $R=0$, in Fig. \ref{fig:vib} we compare the
calculated cross sections with the $\nu_1$ and $2\nu_1$ vibrational
excitation differential cross sections measured at the scattering angles of
30$^\circ$ and 135$^\circ$ \cite{Fabrikant}. Since the ZRP model cross
sections, Eq. (\ref{sigma_n}), are isotropic,
the differential cross section are found as
$d\sigma _{n}/d\Omega =\sigma _n/(4\pi )$. As in Figs. \ref{fig:att} and
\ref{fig:tot}, the two sets of theory curves in Fig. \ref{fig:vib} correspond
to $q_{0}=-0.04$, $\kappa _1=2.0$, and $q_{0}=0.034$, $\kappa _1=4.1$.
The experimental cross sections for the two angles are very different close
to the threshold. However, the data within 30 meV of the threshold may not
be reliable \cite{M_Allan}, while beyond this region the anisotropy of the
differential cross section is greatly reduced. Here the experiment is clearly
in much better agreement with the calculation for a positive value
of $q_0$ (i.e., that for which the anion state is virtual at the equilibrium
of the neutral). This is especially clear for the $n=2$ excitation.
The calculation for $q_{0}=0.034$ and $\kappa _1=4.1$ also shows the same
cusps at higher vibrational thresholds as the experiment.
We thus conclude that the experimental data for the low-energy attachment and
scattering favour the potential curve scheme shown in Fig. \ref{fig:pec}(b).

\subsubsection{Incomplete IVR ($R\neq 0$)}\label{sssec:Rneq0}

In the calculations above we used $R=0$, i.e., we neglected the
reflected wave $\chi ^{(-)}$ in the anion wavefunction (\ref{eq:chi}). In
general, the size of the reflection amplitude $R$ is determined by the rate
of IVR, which depends on the coupling between the vibrational modes. A full
calculation of the vibrational dynamics of highly excited SF$_6^-$ is a
nontrivial task well beyond the scope of the present work.
Hence, we introduce the IVR rate phenomenologically as a width $\gamma $ of the
breathing mode. This is done by adding a small imaginary part $-i\gamma /2$
to the anion potential energy $U_{0}(q)$. The reflection amplitude $R$ is
then obtained semiclassically as outlined in Appendix \ref{app:match}.

In the numerical calculations we treat the anion potential curve for
$q>q_{\rm m}$ in the harmonic approximation. The IVR damping
of the reflected wave is then proportional to $\gamma /\omega _a$,
Eq. (\ref{eq:damp}), where $\omega _a$ is the anion breathing mode frequency.
For strong damping, e.g., $\gamma /\omega _a=1$, the results are
very close those obtained with $R=0$. The reflected wave here is suppressed
by a factor of about $e^{-\pi }\approx 0.05$. In contrast, smaller IVR rates
lead to drastic changes in the cross sections.
In Figs. \ref{fig:att} and \ref{fig:tot} we show the cross sections for
$q_0=0.034$, $\kappa _1=4.1$, and two values of the damping parameter:
$\gamma /\omega _a =0.1$ and 0.01 \cite{comment3}. (The effect of damping
on the cross sections for $q_0<0$ is broadly similar.)

We see that allowing for a sizeable reflected wave results in the emergence
of anion vibrational Feshbach resonances and overall suppression of the
attachment cross section. Both effects spoil the good agreement with experiment
achieved in the calculations with $R=0$. We thus conclude that the IVR in
SF$_6^-$ is very rapid, $\gamma \gtrsim \omega _a$. It takes place over the
time of one vibrational period of the breathing mode, that indicates strong
coupling between the breathing mode and other nuclear degrees of freedom in
SF$_{6}^{-}$.

In the past, the IVR in highly vibrationally excited SF$_6$ molecules was
studied in multiphoton laser excitation experiments (see, e.g., \cite{MP98}
and references therein). These and other studies \cite{A93} indicate that
the IVR rate of the strong infrared-active $\nu _3$ mode is noticeably
smaller than our estimate. However, the two situations are quite different.
In the capture process the amount of energy equal to the electron affinity
of SF$_6$, $E_a=1.06$ eV (recommended value \cite{CO01}), is instantaneously
deposited into a single mode. This mode then has a much higher effective
vibrational quantum number $n\sim E_a/\omega _a \approx 14$, than those in
multiphoton excitation experiments where the energy is distributed
between many modes. At the energies at which it is formed, SF$_6^-$ also has
a stronger anharmonicity related to the proximity of dissociation thresholds.

The vibronic state of SF$_6^-$ at the instant of capture, $\Psi _{\rm att}$,
Eq.~(\ref{NI}), is embedded in the dense multimode vibrational spectrum.
The average level spacing of the SF$_6^-$ vibrational spectrum at the
$e^-+\mbox{SF}_6$ threshold is about $10^{-10}$~a.u.
(see Sec. \ref{sec:decay}). In the process of IVR, the initial state
$\Psi _{\rm att}$ spreads over a large number of multiple vibrational
excitations of SF$_6^-$. For a weak coupling (i.e., small $\gamma $),
$\Psi _{\rm att}$ would describe a simple single-mode vibrational Feshbach
resonance. In the actual case of strong coupling, $\Psi _{\rm att}$ plays
the role of a doorway for the final multimode vibrational resonances.
Therefore, the ultimate states of SF$_6^-$ populated in electron attachment
are extremely closely spaced metastable vibrational resonances with very
large but finite lifetimes. In the next section we consider their decay via
electron autodetachment and estimate the corresponding lifetimes.


\section{Decay of the metastable negative ion}\label{sec:decay}

\subsection{Evaluation of decay widths}\label{subsec:width}

The metastable negative ion species SF$_6^{-*}$ formed by electron
attachment can decay via the reverse, autodetachment process,
Eq.~(\ref{eq:detach}). Another decay channel open at low energies, see, e.g.,
Refs. \cite{Klots1,Lifshitz,Weston}, is dissociation, Eq.~(\ref{eq:dissoc}).
To analyse these possibilities, compare the dissociative attachment cross
section with the
measured cross section for the production of SF$_6^-$ and calculated
$\sigma _{\rm att}$, Fig. \ref{fig:att}. The relatively small SF$_5^-$ signal
at $\varepsilon <150$~meV (below the dissociation threshold) is due to
electron attachment to thermally activated SF$_6$. It depends strongly on the
target molecule vibrational temperature and has been used as a
``thermometer'' \cite{BBM05}. At larger energies the SF$_5^-$ cross section
rises rapidly and at $\varepsilon > 0.3$ eV it becomes the dominant
contribution to the total electron attachment cross section.

In our view this picture indicates that for most of its part dissociation
does not occur as a decay of SF$_6^{-*}$ formed via the resonant doorway state
$\Psi _{\rm att}$. If the latter were true, the calculated cross section
$\sigma _{\rm att}$, which is actually the cross section of formation
of metastable SF$_6^{-*}$, would follow the total attachment cross section,
i.e., the sum of the dissociative attachment cross section and that of
SF$_6^-$ production. We conclude that at low energies, dissociation
does not proceed via the intermediate multimode vibrational
resonances formed in $s$-wave electron attachment described by
$\Psi _{\rm att}$. Hence, autodetachment is the main process responsible
for the decay of SF$_6^{-*}$ \cite{comm:ir}.

\begin{figure}[tbp]
\includegraphics*[width=11cm]{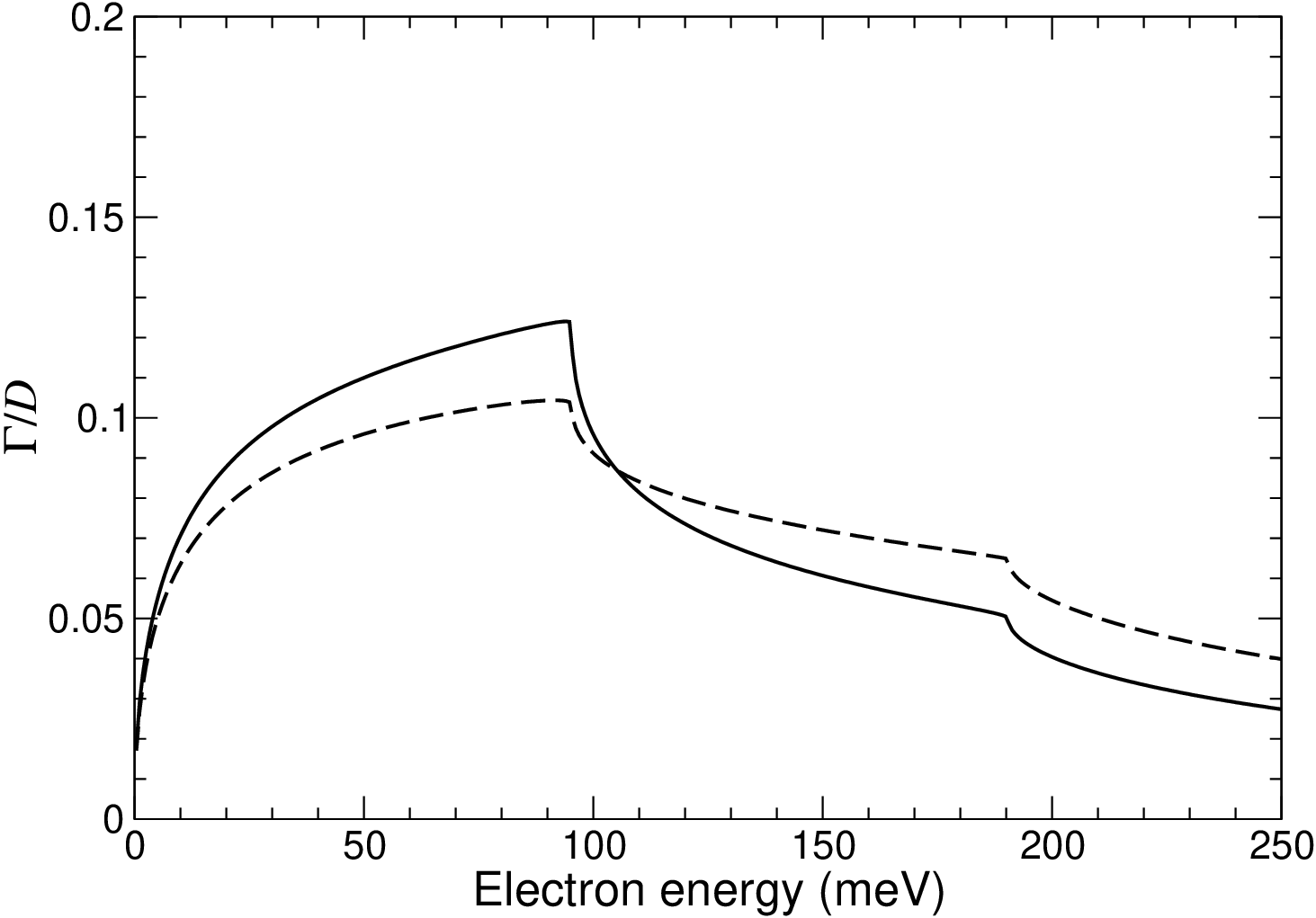}
\caption{Ratio of the mean resonance width to the level spacing evaluated
from the attachment cross section, Eq. (\ref{OM}), for
$q_{0}=0.034$ and $\kappa _1=4.1$, $n_0=0$ (solid curve) and
$n_0=1$ (dashed curve).}
\label{fig:GamD}
\end{figure}

The situation when a projectile forms a very dense spectrum of resonant states
with the target (``compound states'') is well known in nuclear physics
\cite{Bohr}. In this case it is useful to consider the cross section averaged
over an energy interval containing many resonances. Such average cross section
is described by the optical model \cite{Landau}. In this model the resonant
cross section, in our case $\sigma _{\rm att}$, accounts for all processes
occurring via intermediate resonant states (i.e., SF$_6^{-*}$), including
their contribution to elastic scattering. If the mean energy spacing between
the resonances, $D$, is much larger than the resonance widths, the cross
section of electron capture by the SF$_6$ target in the initial vibrational
state $\nu $ can be written as 
\begin{equation}
\sigma _{\rm att}( \varepsilon ,\nu ) =\frac{2\pi ^{2}}{k^2}\,
\frac{\overline{\Gamma }_{\nu }}{D},\label{OM}
\end{equation}
where $\overline{\Gamma }_\nu $ is the average partial elastic width
at a given energy \cite{Landau}. It determines the average
detachment rate leading to a free electron with energy $\varepsilon =k^{2}/2$
and neutral SF$_6$ in state $\nu $. In our approach $\sigma _{\rm att}$
depends only on the number of breathing vibrational quanta in the state $\nu $,
i.e., on $n_0$ (see Sec. \ref{subsec:att}). Figure \ref{fig:GamD} shows the
ratio $\overline{\Gamma }_\nu /D$ obtained from Eq. (\ref{OM}), as a function
of the electron energy for $n_0=0$ and 1.

Since $\overline{\Gamma }_\nu /D\ll 1$, the assumption of non-overlapping
resonances is valid. Hence, we can evaluate $D$ as the reciprocal of the level
density of resonances ($\rho =D^{-1}$), and use Eq. (\ref{OM}) to estimate
the lifetimes of the metastable states, $\tau = \hbar /\Gamma $. In this
context Eq. (\ref{OM}) is sometimes referred to as the principle of detailed
balance \cite{CC66}. In addition to finding $\overline{\Gamma }_\nu $,
one also needs to take into account level-to-level fluctuations of the widths,
which affect the shapes of the SF$_6^{-*}$ decay curves
(see Sec.~\ref{subsec:fluct}).

Apart from the total energy, the process of electron attachment and
detachment considered here conserves the total angular momentum $J$, its
projection $M$, and parity. The electronic part $\psi _{0}$ of the doorway
state
$\Psi _{\rm att}$, Eq. (\ref{NI}), is spherically symmetric, and the continuum
electron is represented by the $s$-wave. Hence, the angular momenta of the
SF$_6$ target and SF$_6^{-*}$ resonances coincide. These resonances also
have the same parity as the initial and final vibrational states of SF$_6$.
Therefore, $D$ in Eq. (\ref{OM}) refers to the average spacing between the
SF$_6^-$ levels with definite $J$, $M$ and parity. The SF$_6$ molecule
is a spherical top, and its rotational states with a given $J$ and $M$ are
$2J+1$ times degenerate with respect to the quantum number $K$ \cite{Landau}.
For the highly-excited SF$_6^-$ this degeneracy can be lifted by ro-vibrational
interactions. The rotational energies of the anion and neutral molecule
are close, and much smaller than the total vibrational energy of SF$_6^-$,
\begin{equation}
E=E_{a}+E_{\nu _i}+\frac{k^{2}}{2},  \label{E}
\end{equation}
where $E_{\nu _i}$ is the target initial vibrational energy. Hence, we
can write
\begin{equation}
D^{-1}=\frac12 (2J+1)\rho (E),
\label{D}
\end{equation}
where $\rho $ is the total density of vibrational states of SF$_6^-$,
and $\frac12$ accounts for parity.

\begin{figure}[tbp]
\includegraphics*[width=11cm]{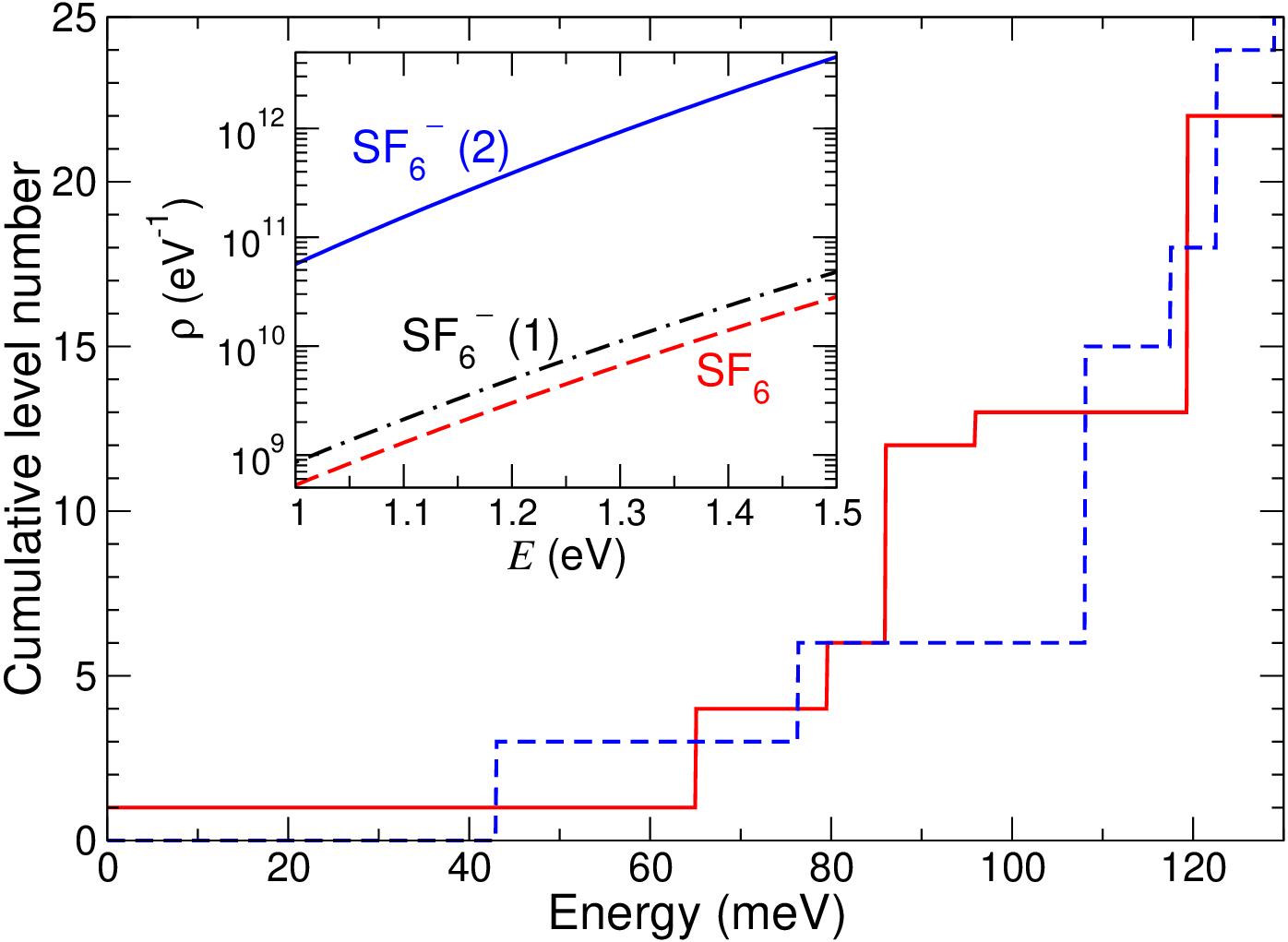}
\caption{Cumulative level numbers of even (solid) and odd (dashed)
vibrational states of SF$_6$. The inset shows the vibrational spectrum
density of SF$_6$ (dashed) and SF$_6^-$ (dot-dashed and solid) obtained using
the frequency sets from Table \ref{tab:freq}.}
\label{fig:dens}
\end{figure}

The total vibrational energy $E$ is large ($\gtrsim 1$~eV), but the mean
energy per vibrational mode is comparable to their frequencies, and we
calculate the vibrational spectrum of SF$_6^-$ using the harmonic approximation
(see Appendix \ref{app:dens}). Table \ref{tab:freq} lists the
mode frequencies of SF$_6$ and SF$_6^-$. In contrast to SF$_6$, the
fundamentals of SF$_6^-$ are not well known. In the past these
frequencies were assumed to be equal to those of the neutral, or slightly
softer \cite{CC66,Klots,Klots1,Lifshitz,Weston}. For example, the first set of
anion frequencies in Table \ref{tab:freq} was used in
Refs. \cite{Lifshitz,Heneghan}. The second set is from Ref. \cite{GB98},
which is probably the best calculation of SF$_6$ and SF$_6^-$ by the
coupled-cluster and many-body perturbation theory (MBPT) methods. These
lower anion frequencies are supported by a recent photodetachment
study of SF$_6^-$ \cite{BR07}, and we regard them as more accurate.
On the inset of Fig. \ref{fig:dens} we show the total vibrational level
densities in the energy range relevant to metastable SF$_6^{-*}$.
As expected, the density obtained using the second set of anion
frequencies is much greater than the density given by the first
set.

\begin{table}[tbp]
\caption{Vibrational modes and frequencies of SF$_6$ and SF$_6^-$.}
\label{tab:freq}
\begin{ruledtabular}
\begin{tabular}{ccccc}
\hline
Mode & Symmetry & \multicolumn{3}{c}{Frequencies (cm$^{-1}$)} \\
&&SF$_6$\footnotemark[1]&SF$_6^-$\footnotemark[2]&SF$_6^-$\footnotemark[3] \\
\hline
1 & $A_{1g}$ & 774 & 700 & 626 \\
2 & $E_{g}$  & 642 & 625 & 447 \\
3 & $T_{1u}$ & 948 & 925 & 722 \\
4 & $T_{1u}$ & 616 & 594 & 306 \\
5 & $T_{2g}$ & 525 & 500 & 336 \\
6 & $T_{2u}$ & 347 & 325 & 237 \\
\end{tabular}
\footnotetext[1]{Experimental data from Ref. \cite{Shim}.}
\footnotetext[2]{Frequencies used in Ref. \cite{Lifshitz} and earlier
in Ref. \cite{Heneghan}.}
\footnotetext[3]{MBPT(2) calculation, Ref. \cite{GB98}.}
\end{ruledtabular}
\end{table}

The average total resonance width $\Gamma $ is the sum of partial widths
$\overline{\Gamma }_\nu $ for all open decay channels $\nu $ allowed by the
conservation of energy, angular momentum and parity. Equations (\ref{OM})
and (\ref{D}) give
\begin{equation}
\Gamma (E)=\sum_{\nu }\frac{\sigma _{\rm att}\left( \varepsilon _{\nu },\nu
\right) k_{\nu }^{2}}{\pi ^{2}\rho (E)},  \label{Klots}
\end{equation}
where the energy $\varepsilon _{\nu }$ and momentum $k_{\nu }$ of the
emitted electron are determined by $\varepsilon _\nu =k_\nu ^2/2=
k^2/2+E_{\nu _i}-E_\nu $. The sum in Eq. (\ref{Klots}) is over the
open-channel vibrational states of neutral SF$_6$. Summation over $2J+1$
states with different $K$ gives an additional factor $2J+1$, which is
cancelled by the same factor in Eq. (\ref{D}). To compare with experiment,
the width (\ref{Klots}) can also be averaged over the distribution of the
initial target states and incident electron energies.

To within a factor of two related to parity selection, Eq. (\ref{Klots})
coincides with that derived by Klots \cite{Klots} using equilibrium
considerations. It has a simple physical meaning. The decay rate in each
channel is given by the probability per unit time for SF$_6^{-*}$ to enter
the doorway state $\Psi _{\rm att}$, which is equal to the classical frequency
$D/2\pi $, times the probability $P$ of detachment at each attempt.
The latter quantity is determined by the capture cross section,
$P=\sigma _{\rm att}\left( \varepsilon ,\nu \right) k^{2}/\pi $,
as follows from the detailed balance relations \cite{CC66}.

\begin{figure}[tbp]
\includegraphics*[width=11cm]{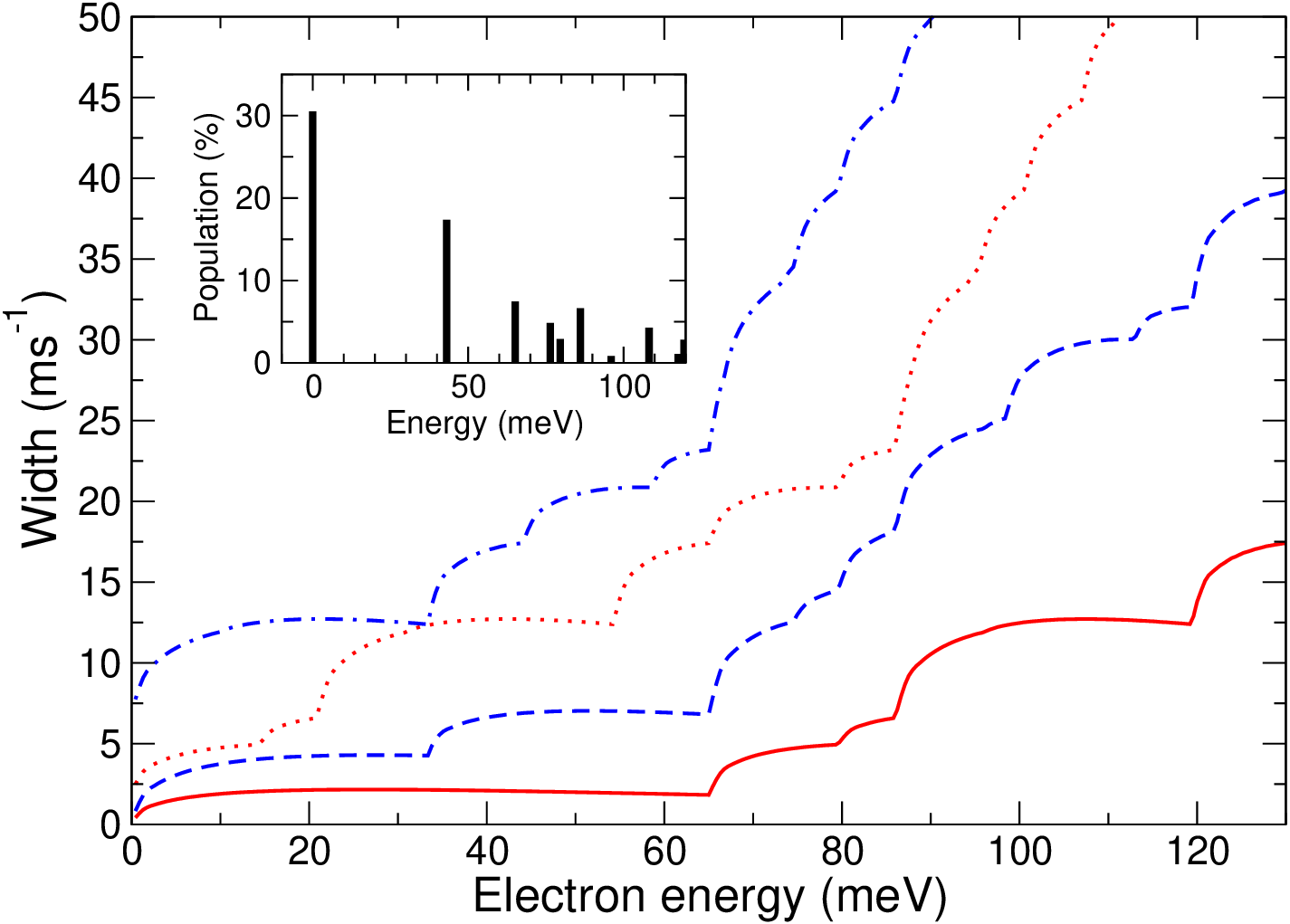}
\caption{Mean resonance width $\Gamma $ as a function of the incident electron
energy $\varepsilon $ for different initial target states: ground state (solid
curve), $T_{2u}$ (dashed curve), $T_{2g}$ (dotted curve), and $T_{1u}$
(dot-dashed line). Shown on the inset are the populations of a few lower
vibrational states of SF$_6$ at room temperature.}
\label{fig:Gam}
\end{figure}

To show how the number of open decay channels affects the width, we plot in
Fig. \ref{fig:dens} the cumulative number of SF$_6$ vibrational excitations
for both parities. The parity of the final SF$_6$ must be the same as that
of its initial state. Besides the ground state, at room temperature
only the lowest excited states, $T_{2u}$, $T_{2g}$, and $T_{1u}$, are
populated with probabilities $w_{\nu }>0.05$. These are shown on the inset of
Fig. \ref{fig:Gam}. In this figure we also show the energy dependence
of the mean resonance width $\Gamma $, Eq. (\ref{Klots}), for each
of the above target states. The dependence of $\Gamma $ on
$\varepsilon $ is step-like, each step related to the opening of a new decay
channel. The width is also bigger for higher-lying target states,
as more channels are open.

The widths in Fig. \ref{fig:Gam} are given in reciprocal time units,
i.e., they represent the decay rates $\Gamma /\hbar $. The smallest of the
widths, for the electron capture by the ground-state SF$_6$, is about
2 ms$^{-1}$ below 65 meV. This indicates a lifetime of 0.5 ms, which is
comparable to the values observed in traps \cite{Henis,Odom,Dunning,Dunning1}.
Note however, that the widths are very sensitive to the
SF$_6^-$ vibrational spectrum.
Thus, if we used the first set of anion frequencies from Table~\ref{tab:freq},
the widths given by Eq. (\ref{Klots}) would be about $10^2$ times larger,
since they are inversely proportional to the vibrational level density of
SF$_6^-$ shown in Fig.~\ref{fig:dens}. The smallest width would then be about
10~$\mu $s, i.e., in the range of values from time-of-flight experiments
\cite{EG62,CC66,HT71,DA86,AD88,Garrec}. In fact, the earliest of such
studies, Ref. \cite{CC66}, which analysed the lifetimes theoretically
in a way similar to Eq. (\ref{Klots}), using SF$_6$ fundamentals and
disregarding parity, concluded that the lifetime of 25~$\mu$s corresponded
to the electron affinity of $\sim 1.1$ eV.

It should also be noted that due to the strong energy dependence of
$\rho (E)$, the widths and lifetimes are sensitive to the value of the electron
affinity. A change of 0.1~eV in $E_a$ changes the lifetimes
by a factor of three. In addition, our calculation of the density neglects
anharmonic effects in the anion vibrational spectrum.
The amount of energy deposited in each mode is relatively small, but the
total vibrational energy $E$ is close to the dissociation threshold. In this
case anharmonicity can increase $\rho (E)$, and hence the lifetimes, by
a factor of two or three.

Figure \ref{fig:Gam} shows that if the electrons and SF$_6$ molecules
possess broad thermal energy distributions, e.g., when the anions are
formed in a trap, the decay of SF$_6^{-*}$ will be characterised
by a set of average widths rather than a single detachment width. This was
indeed observed by Delmore and Appelhans~\cite{DA86,AD88},
who detected several lifetimes or bunches of lifetimes depending on the
temperature (in the 10~$\mu $s range, though). For monoenergetic electrons
with $\varepsilon <65$~meV and vibrationally cold SF$_6$, the decay process
is governed by a single average width. Such experiment was done by
Garrec {\em et al.} \cite{Garrec}, who did report a single lifetime
$\tau \approx 19~\mu $s. However, the corresponding width would be close to
the values found using the neutral or similar vibrational
frequencies~\cite{Lifshitz}. It is incompatible with the more accurate
softer anion vibrational spectrum, unless a much lower electron affinity
is used.

In our view, the present calculation does explain observations of sizeable
SF$_6^{-*}$ signals at millisecond and greater times, and its slow
nonexponential decay \cite{Odom,Dunning,Dunning1}, at least qualitatively.
Equations (\ref{OM}) and (\ref{Klots}) yield the widths
{\em averaged} over large numbers of closely spaced resonances. Given the
large density $\rho (E)$, these resonances cannot be resolved experimentally,
even with a highly monoenergetic electron beam. As a result, one always
observes an ensemble of such states. Some of them may have widths much
smaller than the average. Such states will form a tail of long-lived
anions, as was first suggested by Odom~{\em et al.}~\cite{Odom}.
Therefore, to determine the survival curve of SF$_6^{-*}$, one must take
into account the {\em distribution} of resonance widths.

\subsection{Fluctuations of the widths and nonexponential decay}
\label{subsec:fluct}

The width of a particular multimode vibrational resonance is determined by
the size of the doorway state $\Psi _{\rm att}$ component in its
wavefunction. This component is extremely small since $\Psi _{\rm att}$
spreads over a large number of such resonances, which can be estimated
as $\gamma \rho (E)\sim \omega _a\rho (E)\sim 10^{10}$
(Sec. \ref{sssec:Rneq0}). Physically, this situation is similar to that of
neutron capture by heavy nuclei, where each of the narrow compound resonances
contains only a small fraction of the ``neutron $+$ target'' state, which
allows their coupling to the continuum. Due to strong mixing between the
basis states that make up the compound states, the statistics of their
components becomes Gaussian (see, e.g., Ref. \cite{Brody}). This leads to
the following probability density for the partial widths $\Gamma _\nu$,
\begin{equation}\label{PT}
f(\Gamma _\nu )=\frac{e^{-\Gamma _\nu /2\overline{ \Gamma }_\nu }}
{\sqrt{2\pi \Gamma _\nu \overline{\Gamma }_\nu }},
\end{equation}
where $\overline{\Gamma }_\nu $ is the mean.
Equation (\ref{PT}) is known as the Porter-Thomas distribution
\cite{Bohr,PT56}.

If only one decay channel is open, every resonance decays exponentially as
$e^{-\Gamma _\nu t}$. However, it is easy to see that fluctuations
of $\Gamma _\nu$ result in a nonexponential time dependence of the survival
probability $P(t)$ for the ensemble. Assuming that at $t=0$ all levels in the
ensemble have equal populations and using Eq. (\ref{PT}), one obtains,
\begin{equation}\label{eq:Pt}
P(t)=\int _0^\infty e^{-\Gamma _\nu t}f(\Gamma _\nu )d\Gamma _\nu
=(1+2\overline{\Gamma} _\nu t)^{-1/2}.
\end{equation}
For more than one decay channel, the survival probability
is given by (see, e.g., \cite{Miller,KA00})
\begin{equation}\label{eq:PtN}
P(t)=\prod _\nu (1+2\overline{\Gamma} _\nu t)^{-1/2},
\end{equation}
where the product includes all open channels.
When the number of open channels, $N_c$, is large, then for
$\overline{\Gamma} _\nu t\ll 1$ one can use
$(1+2\overline{\Gamma} _\nu t)^{-1/2}\simeq e^{-\overline{\Gamma }_\nu t}$
in Eq. (\ref{eq:PtN}), which gives
\begin{equation}\label{eq:exp}
P(t)\simeq e^{-\Gamma t},
\end{equation}
where
$\Gamma =\sum _\nu \overline {\Gamma }_\nu \sim N_c\overline{\Gamma }_\nu$
is the total width. The exponential behaviour that holds for
$\Gamma t<\Gamma /\overline{\Gamma} _\nu \sim N_c$ is a consequence
of suppression of fluctuations in the total width. The case of $N_c\gg 1$
in fact corresponds to the classical limit of the decay process
of a compound system with a dense spectrum of states.
The parameter $N_c$ determines whether the decay process is
classical with an exponential behavior, or quantum where sizeable
deviations from the pure exponent can be expected (see, e.g.,
\cite{Cavity}).

Besides the fluctuations of the partial widths described by Eq. (\ref{PT}),
the distribution of resonance widths in the negative ion ensemble depends
on the conditions of its formation. Thus, SF$_6^{-*}$ formed in a
trap will have a different distribution of autodetachment widths
and hence, different decay curves, to SF$_6^{-*}$ formed with a
high-resolution electron beam. In particular, the population of SF$_6^-$
resonances created in a beam experiment at time $t=0$ is proportional to
the level density and their elastic (entrance) widths. For a single decay
channel $\nu $ (identical to the entrance channel), the number of ions
that have survived to time $t$ is found as
\begin{equation}\label{eq:Nt}
N(t)\propto \rho (E)\int \Gamma _\nu e^{-\Gamma _\nu t}
f(\Gamma _\nu )d\Gamma _\nu =\frac{\overline{\Gamma} _\nu \rho (E)}
{(1+2\overline{\Gamma} _\nu t)^{3/2}}.
\end{equation}
Note that compared with Eq. (\ref{eq:Pt}), the account of the initial resonance
population has resulted in an additional factor
$(1+2\overline{\Gamma} _\nu t)^{-1}$. For more than one decay channel,
Eq. (\ref{eq:PtN}) will be similarly modified,
\begin{equation}\label{eq:NtNJ}
N(t)\propto \frac{\overline{\Gamma} _{\nu _i}\rho (E)}
{1+2\overline{\Gamma} _{\nu _i}t}
\prod _\nu (1+2\overline{\Gamma} _\nu t)^{-1/2},
\end{equation}
where $\nu _i$ is the initial vibrational state of the target molecule.

If the target molecules are characterised by a distribution of initial
states with vibrational and rotational quantum numbers $\nu _i$ and $J$,
with probabilities $w_{\nu _i J}$, the decay curves must be averaged
over these. This gives
\begin{equation}\label{eq:Nt_av}
N(t)\propto \sum _{\nu _i\,J}\frac{w_{\nu _iJ}\overline{\Gamma} _{\nu _i}
\rho (E)}{1+2\overline{\Gamma} _{\nu _i}t/N_J}
\prod _\nu (1+2\overline{\Gamma} _\nu t/N_J)^{-N_J/2},
\end{equation}
where $N_J$ is the number of channels with different rotational quantum
numbers $K'$ of the final SF$_6$ that can be populated for a given
$K$ of the target molecule. If the $K$ quantum number is not conserved than
$N_J=2J+1$, while if $K$ were conserved, one would have $N_J=1$. In the
former case the decay should be close to exponential for room temperature
(or even much colder) SF$_6$, since typical $J$ and $N_J$ are large.
Indeed, the probabilities for a thermal ensemble of SF$_6$ molecules
are $w_{\nu _iJ}\propto \exp \{-[E_{\nu _i}+J(J+1)/2I]/T\}$, where
$T$ is the temperature in energy units, and $I=1.23\times 10^6$ a.u. is the
moment of inertia of SF$_6$. Hence, from $J(J+1)/2I \sim \frac 32 T$ we have
$J\sim 50$ at room temperature.

The widths in Eq. (\ref{eq:Nt_av}) depend on the initial vibrational state
of the target and the incident electron energy. For SF$_6^{-*}$ formed in a
trap in equilibrium with thermal electrons, the resonances will be
populated uniformly, with Boltzmann probabilities $\propto e^{-E/T}$.
In this case the decay curve is given by the right-hand side of
Eq. (\ref{eq:Nt_av}) without the factor
$w_{\nu _iJ}\overline{\Gamma}_{\nu _i}/(1+2\overline{\Gamma}_{\nu _i}t/N_J)$,
averaged over the Boltzmann anion energy distribution.

\begin{figure}[tbp]
\includegraphics*[width=11cm]{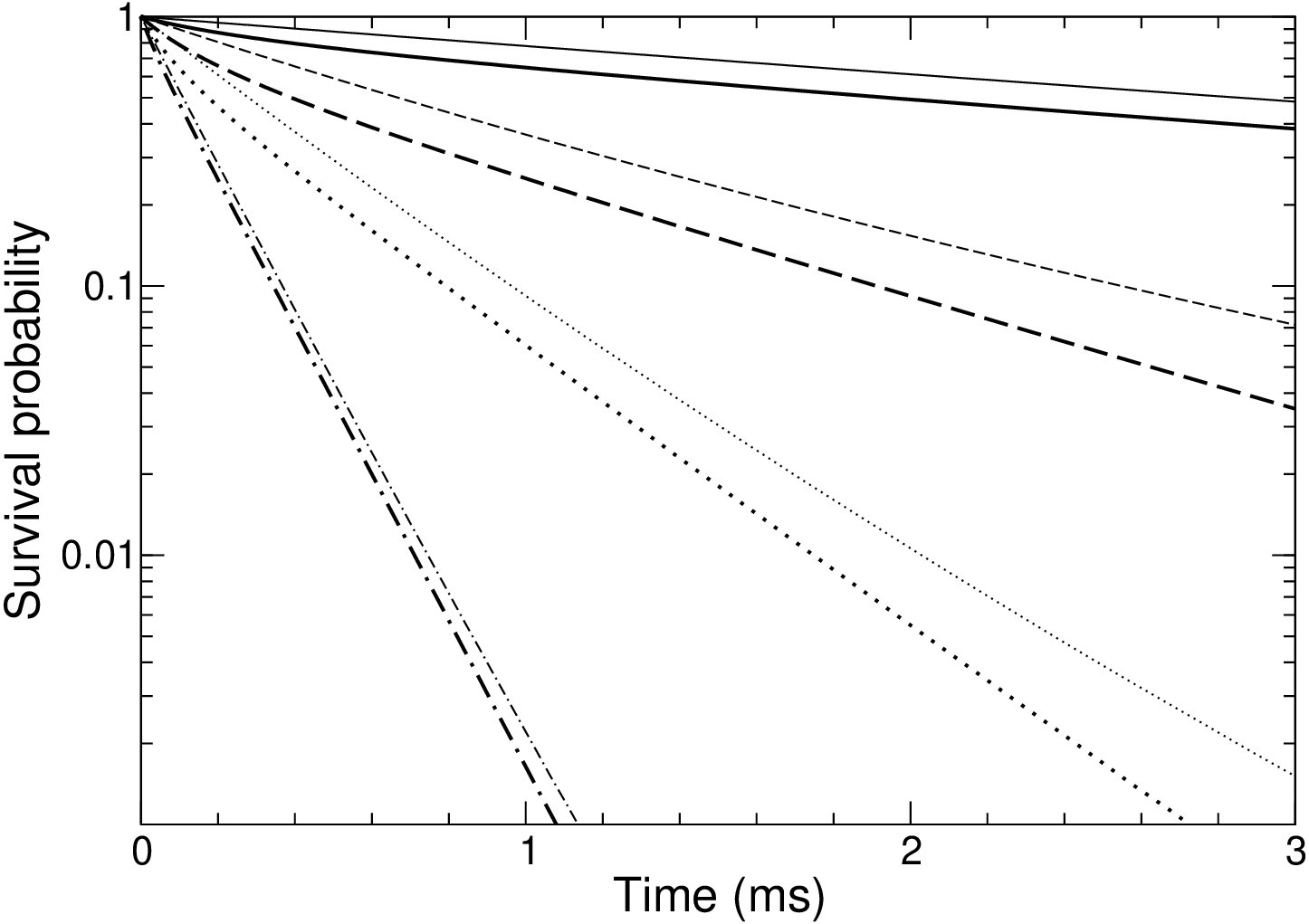}
\includegraphics*[width=11cm]{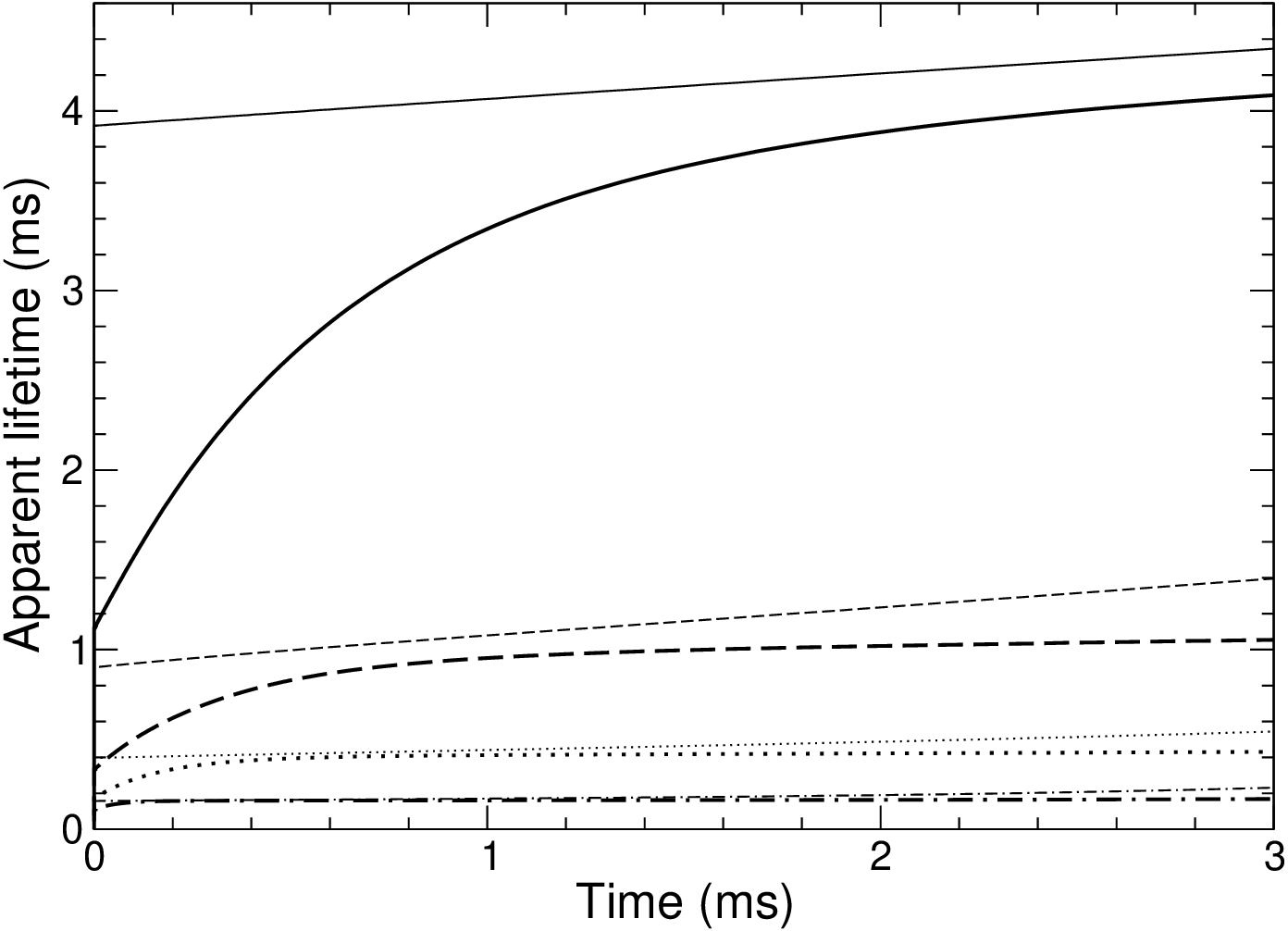}
\caption{Time evolution of the survival probability of SF$_6^{-*}$, $N(t)/N(0)$
(top) and apparent lifetimes, $\tau $ (bottom) calculated for two target
temperatures $T=300$ K (thick curves) and $T=10$ K (thin curves), and
different electron energies: 0.45 meV (solid), 44 meV (dashed), 79 meV
(dotted), and 113 meV (dot-dashed).}
\label{fig:decay12}
\end{figure}

\begin{figure}[tbp]
\includegraphics*[width=11cm]{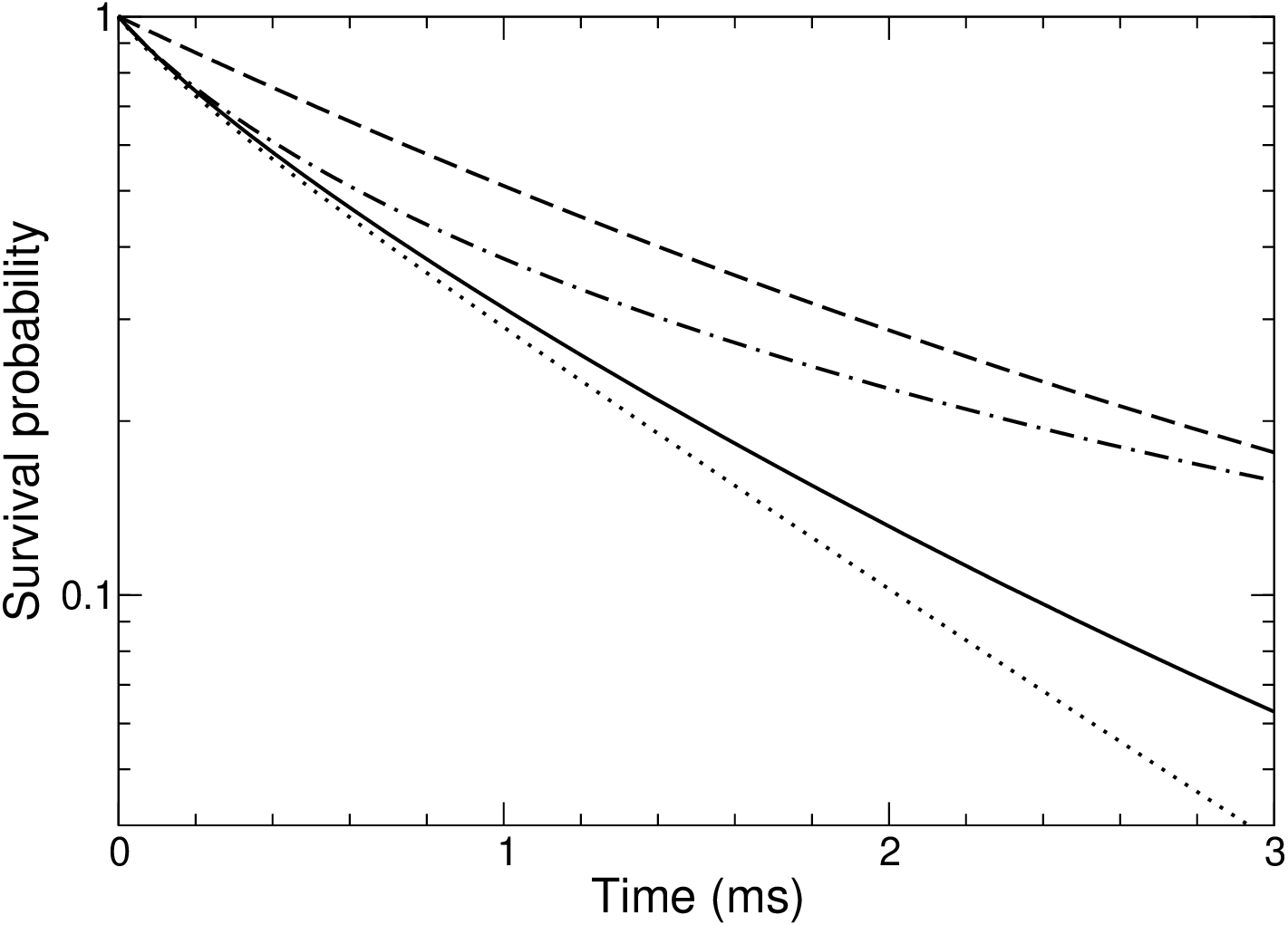}
\includegraphics*[width=11cm]{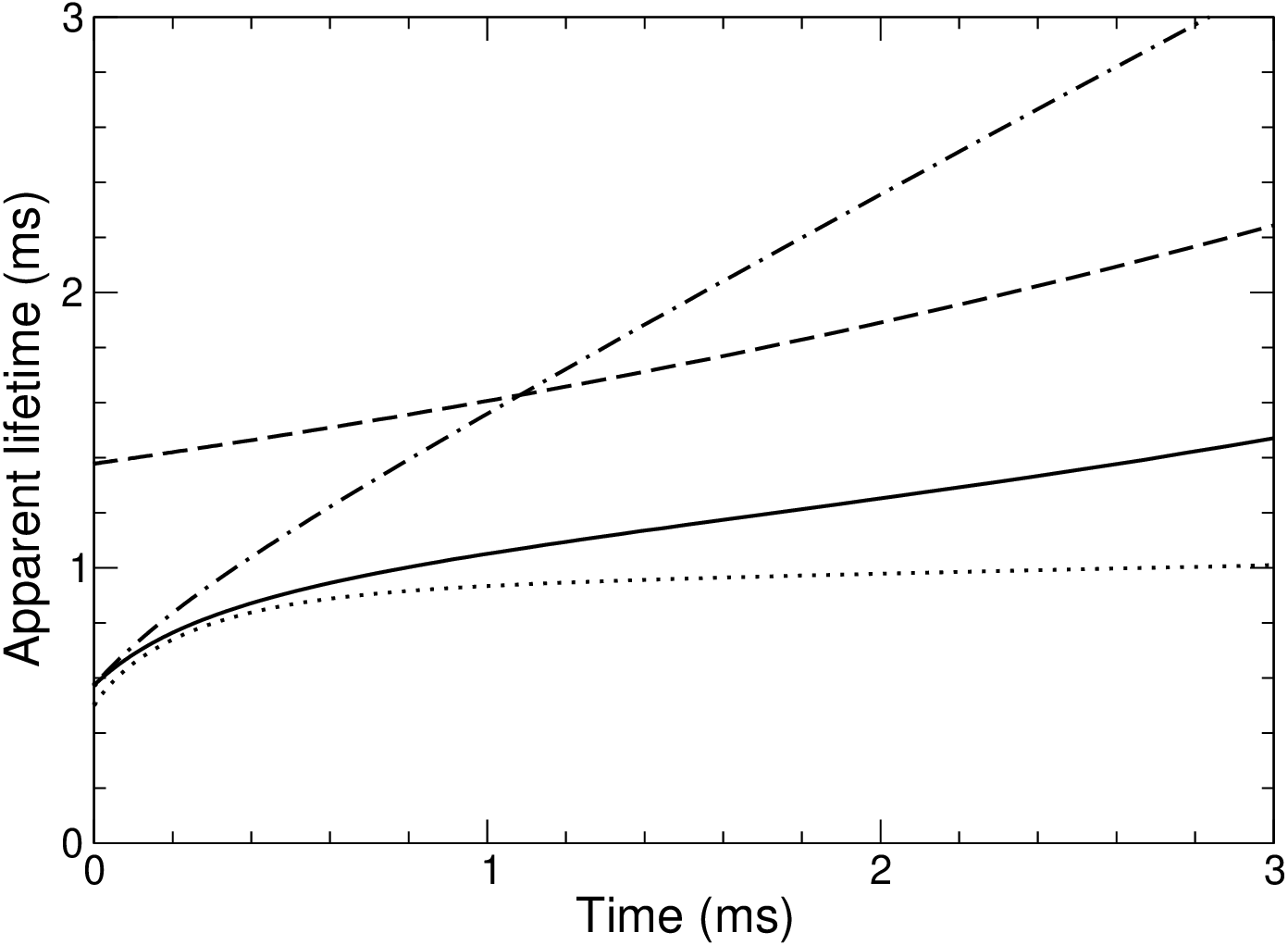}
\caption{Time evolution of the survival probability of SF$_6^{-*}$,
$N(t)/N(0)$ (top) and apparent lifetimes, $\tau $ (bottom), calculated for
anions formed at equilibrium conditions at two trap temperatures, $T=300$ K
(solid) and $T=77$ K (dashed). Dotted curve corresponds to the decay of
SF$_6^{-*}$ produced by capture of electrons with fixed energy
$\varepsilon = 38$ meV, equal to the mean energy at room temperature.
Dot-dashed curve shows $\tau $ obtained assuming conservation of the angular
quantum number $K$ (i.e., for $N_J=1$).}
\label{fig:decay34}
\end{figure}

Figures \ref{fig:decay12} and \ref{fig:decay34} show the results of
our modelling of the decay curves of SF$_6^{-*}$ formed under
different conditions. In both figures the bottom plates show the apparent
lifetime $\tau $ estimated from the instantaneous decay rate
$\tau ^{-1}=(dN/dt)/N$. This quantity should be constant for a purely
exponential decay, but in general, given Eqs. (\ref{eq:Nt})--(\ref{eq:Nt_av}),
increases with time. The rate of such increase is asymptotically linear,
$\tau \propto t$, since the fraction of long-lived ions increases with time.

In Fig. \ref{fig:decay12} we assume that the anions are produced in collisions
of monochromatic electrons with four different energies with the SF$_6$ gas at
$T=300$ and 10 K. As expected from the energy dependence of the widths,
Fig. \ref{fig:Gam}, the decay becomes faster much with the increase of the
electron energy. However, even at the highest energy the lifetime remains
greater than 100 $\mu$s, which is compatible with the laser photoelectron
attachment experiments \cite{KRH92,BBM05}.
Non-exponential features clearly seen in Fig. \ref{fig:decay12} at small
times are due to the contribution of short-lived anions formed by
vibrationally excited target molecules. Naturally, this effect is more
pronounced at higher temperatures. The nonexponential decay at
large times is caused by the Porter-Thomas fluctuations of the autodetachment
widths. This effect is greater at small molecular temperatures, where
the number of rotational channels is not too large. However, at room
temperature $N_J\sim 50$, and the decay is close to exponential over the
time interval shown.

The curves that model the decay of SF$_6^{-*}$ formed in a trap at two
temperatures, $T=300$ and 77~K, are shown in Fig. \ref{fig:decay34}.
In addition to nonexponential features at small times, the presence of
long-lived anions formed by low-energy electrons leads to a nonexponential
decay at large times. To illustrate this effect we show for comparison in
Fig. \ref{fig:decay34} the decay curve for anions formed by monoenergetic
electrons with the energy of $\frac 32 T$. Also shown in
Fig.~\ref{fig:decay34} is the decay curve calculated assuming that $K$ is
conserved (i.e., neglecting the effect of the rotational motion on the number
of decay channels). As expected, this decay curve is strongly nonexponential,
with the apparent lifetime growing as $\tau \sim t$. This feature of the
SF$_6^-$ decay was observed experimentally in Refs.~\cite{Odom,Dunning1}.
It is a consequence of the fact that the fraction of surviving long-lived
ions increases with $t$. On the other hand, allowing for the mixing of the
rotational quantum number $K$ makes for a faster (exponential) depletion of
SF$_6^-$, which makes it hard to explain how the anion signal can be
observed at much longer times.

There is another effect that molecular rotations can have on the SF$_6^{-*}$
lifetimes, and that has not been taken into account in the present
consideration. Due to a difference between the equilibrium S--F bond lengths
of the neutral and anion, the moment of inertia of SF$_6^-$ is about 20\%
greater than that of the neutral. This means that the contribution
of the rotational energy $J(J+1)/2I$ in SF$_6$ and SF$_6^-$ differs by
20\% as well. At room temperatures this difference is close to 10 meV,
which means that the amount of energy available for IVR in different
rotational states is different, and is larger for higher $J$.
This is another source of fluctuations in the detachment widths, which
can in principle affect the anion decay curves.

\section{Conclusions}
Comparison of calculated cross sections with experimental data demonstrates
that low-energy electron attachment to SF$_6$ proceeds via capture into a
virtual state with a strong coupling to the breathing vibrational mode. The
model provides an accurate description of the attachment cross section, and
is in agreement with the measured total and vibrational excitation cross
sections. From this comparison, we have determined the two parameters of the
model which describe the behaviour of the SF$_6$ and SF$_6^-$ potential curves
near the equilibrium of the neutral. This behaviour is in accord with the
quantum chemistry calculations of the potential curves.

By allowing a reflected wave in the nuclear dynamics of the SF$_6$ breathing
mode, we have studied the effect of the IVR rate on the cross sections.
Comparison with experiment indicates that the IVR in SF$_6^-$ formed
by electron attachment is very fast, its rate being comparable to the
frequency of the breathing mode.

We have evaluated the autodetachment widths of metastable SF$_6^{-*}$
resonances, assuming statistical distribution of the energy over the
vibrational spectrum of the molecule. The magnitude of the widths depends 
strongly on the set of SF$_6^-$ vibrational frequencies used, as well as
on the adiabatic electron affinity of SF$_6$. Using the recommended value
of 1.06 eV together with the best calculated fundamentals we obtain
estimates of SF$_6^{-*}$ lifetimes in the 100 $\mu $s to 1 ms range,
shorter lifetimes corresponding to the anions formed by higher-energy
electrons. These lifetimes are broadly in agreement with the values
inferred in laser photoelectron attachment experiments and found
in ion-cyclotron-resonance experiments and in traps. At the same time,
we cannot explain the observation of tens of $\mu $s lifetimes in
time-of-flight experiments. Such values would be compatible with a much
stiffer set of SF$_6^-$ fundamentals (similar to the neutral), or smaller
electron affinity values.

By using the Porter-Thomas distribution we have modelled the effect of
fluctuations of autodetachment widths due to statistical (``chaotic'')
nature of highly-excited vibrational states of SF$_6^-$. 
We have also investigated the effect of SF$_6$ temperature and electron
energy on the SF$_6^{-*}$ decay curves. Fluctuations of the widths 
result in nonexponential decay of SF$_6^{-*}$. However, the presence
of a large number of rotational channels for all but very low temperatures
of the neutral molecule makes these effects relatively small.

Finally, dissociative attachment into $\mbox{SF}_5^- + \mbox{F}$ has been
largely ignored in the present work. We believe that it can have only a
very small effect on the lifetimes of metastable SF$_6^-$ formed at
low electron energies. It appears that for electron energies below 0.2 eV
the $\mbox{SF}_5^-$ ions originate from (thermally activated) metastable
SF$_6^-$. However, the dissociation signal is low due to a
small branching ratio in comparison with autodetachment. At higher electron
energies the SF$_5^-$ signal has a large peak. Here the dissociation cross
section exceeds that of the $s$-wave attachment model. In our view this
means that the main dissociation mechanism at these energies is different
from that responsible for the long-lived SF$_6^-$.

When this work was completed, we became aware of two very recent studies 
of the electron-SF$_6$ problem. In the first of these \cite{CLS07}, the
autodetachment
lifetimes of metastable SF$_6^-$ were studied with a time-of-flight and
Penning ion trap techniques, for very low ($\sim 1$~meV) electron energies
as a function of the gas temperature. At room temperatures only long-lived
SF$_6^{-*}$ with $\tau \gtrsim 1 $~ms were seen. At higher temperatures
shorter lifetimes were observed ($\sim 0.4$ ms), together with a small signal
of short-lived anions ($\lesssim 10$~$\mu $s). These values are broadly in
agreement with our decay curve modelling. The second set of papers
\cite{TMV07} used kinetic modelling within the framework of the statistical
unimolecular rate theory. It analysed the attachment and dissociation
data in a wide range of target and electron temperatures, as well as SF$_6$
and carrier gas pressures, allowing for additional effects (e.g., the rate
of IVR as a function of the incident electron energy) by model factors and
fitting parameters. One of the results of this analysis is an indication
of a larger electron affinity of $1.20\pm 0.05$~eV. This value would
result in a factor of 3--5 decrease of our autodetachment widths, and a
corresponding increase of the lifetimes. Given the sensitivity of the decay
curves to the conditions under which SF$_6^{-*}$ are formed, such change
still leaves the lifetimes in the 1--10 ms range, for the conditions
similar to those used in Ref. \cite{CLS07}.

\begin{acknowledgments}
We thank M. Allan, D. Field and H. Hotop for providing their experimental
data in numerical form, and are grateful to M. Allan, F. B. Dunning,
T. Field, H. Hotop and T. M. Miller for useful discussions. This work has
been supported by the International Research Centre for Experimental
Physics (Queen's University Belfast).
\end{acknowledgments}


\appendix

\section{Matching of the wavefunction}\label{app:match}

Here we show how the total wavefunction $\Psi ({\bf r},q)$ of the
$e^-+\mbox{SF}_6$ system, Eq. (\ref{eq:WF1}), with the amplitudes satisfying
Eq. (\ref{eq:f_n}), matches the adiabatic wavefunction
$\Psi _{\rm att}({\bf r},q)$, Eq. (\ref{NI}), of SF$_6^-$, in the range of
nuclear coordinates 
where the incident electron is bound. In order to do this, we project
$\Psi _{\rm att}$ onto the SF$_6$ breathing vibrational states $\chi _{n}(q)$ 
with large $n$. This allows us to determine the asymptotic behaviour of $f_n$
corresponding to the two terms on the right-hand side of Eq. (\ref{eq:chi})
and obtain an explicit relation between $f_n$ and the amplitudes $A$ and $B$.

Consider the wavefunction $\Psi _{\rm att}({\bf r},q)$ far from the merging
point of the neutral and anion potential curves. Here its nuclear part
(\ref{eq:chi}) can be written explicitly using the semiclassical (WKB)
approximation \cite{Landau}. Let us first consider the contribution of
the outgoing wave term,
\begin{equation}\label{eq:chiDA}
\chi (q)=\frac{A}{\sqrt{v(q)}}\exp \left( i\int _a^q p(q)dq\right) ,
\end{equation}
where $p(q)=\sqrt{2M[E-U_0(q)]}$, $v(q)=p(q)/M$ is the corresponding
velocity,
$E=\frac 12 k^2+E_{n_0}$ is the energy of the system, and $a$ is in the
classically allowed region (its choice only affecting the phase of $A$).

Let us now demonstrate that the projection of
$\Psi _{\rm att}=\psi _{0}({\bf r},q)\chi (q)$ onto
$\chi _{n}(q)$,
\begin{equation}\label{eq:int1}
g_n({\bf r})=\int \Psi _{\rm att}({\bf r},q) \chi _{n}(q) dq,
\end{equation}
matches the electronic part of the terms in the sum over $n$ in
Eq. (\ref{eq:WF1}).

For a large $n$, we can use a normalised semiclassical
expression for $\chi _{n}(q)$ \cite{comment},
\begin{equation}\label{eq:chi_n}
\chi _{n}(q)=\sqrt{\frac{2\omega }{\pi v_{n}(q)}}\cos \left(
\int_{q}^{a_n}p_{n}(q)dq-\frac{\pi }{4}\right) ,
\end{equation}
where $p_{n}=\sqrt{2ME_n-(M\omega q)^2}$, $v_n(q)=p_n(q)/M$,
and $a_n=\sqrt{2E_n/M\omega ^2}$ is the classical turning point. Using
Eqs. (\ref{eq:chiDA}) and (\ref{eq:chi_n}) in (\ref{eq:int1}), we obtain:
\begin{equation}\label{eq:int2}
g_n({\bf r})=\frac{A\sqrt{\omega }}{\pi r}\int
\frac{\sqrt{\kappa (q)}e^{-\kappa (q)r}}{\sqrt{v(q)v_n(q)}}
\exp \left( i\int _a^q p(q)dq \right) \cos \left(
\int_{q}^{a_n}p_{n}(q)dq-\frac{\pi }{4}\right) dq.
\end{equation}
The integral above contains rapidly oscillating functions. It can be
evaluated in the saddle-point approximation, and the main contribution
is due to the incoming wave component of the cosine. Hence, the
oscillatory factor in the integrand is of the form
$\exp [i\varphi _n(q)]$, where
\begin{equation}\label{phi}
\varphi _n(q)=\int _a^{q} p(q)dq +\int_q^{a_n}p_n(q)dq .
\end{equation}
The equation for the saddle point, $\partial \varphi _n/\partial q =0$, gives
$p(q)=p_n(q)$, i.e., the ``transition'' between the nuclear vibrational
state of the neutral and that of the anion takes place when the momenta
are equal. This gives the following equation for the saddle point $q_n$:
\begin{equation}\label{kappa-n}
E-\varepsilon _{0}(q_{n})=E_n .
\end{equation}
It immediately follows that
\begin{equation}\label{qn}
\kappa (q_{n})=\sqrt{-2\varepsilon (q_n)}=\sqrt {2(n-n_{0})\omega -k^2}
\equiv |k_n|,
\end{equation}
which shows that $g_n({\bf r})\propto e^{-|k_n|r}/r$, exactly as that of
the $n$th term in Eq. (\ref{eq:WF1}) for closed channels.
Note also that the saddle point, found using Eq. (\ref{eq:chi_n}), as
\begin{equation}\label{eq:q_n}
q_n=\frac{|k_n|-\kappa _0}{\kappa _1}
\sim \frac{\sqrt{2\omega n}}{\kappa _1},
\end{equation}
lies in the classically allowed region of the oscillator, which justifies
the use of Eq. (\ref{eq:chi_n}).

Completing the saddle-point calculation of the integral (\ref{eq:int2}),
we obtain an expression for the amplitude $f_n$ at large $n$, for the case
when $\chi (q)$ is an outgoing wave:
\begin{eqnarray}\label{fn1}
f_n &=& A\sqrt{\frac{\omega }{2\pi v(q_n)\kappa _1}}
\exp [ i\varphi _n(q_n)]\\
&\simeq &A\sqrt{\frac{\omega }{2\pi \kappa _1}}
\left[\frac{2\omega n}{M}\left( 1-\frac{M\omega ^2}{\kappa_1^2}
\right)\right]^{-1/4}\exp [ i\varphi _n(q_n)] .\label{eq:fna}
\end{eqnarray}
In the last equation we used an asymptotic expression for the velocity
$v(q_n)=\sqrt{2[E-U_0(q_n)]/M}$ obtained for large $n$ using
Eq. (\ref{eq:U_0}).

Thus we see that apart from a slowly varying pre-factor, the successive
amplitudes differ by their phase, so that
\begin{equation}\label{eq:fn1fh}
\xi =\frac{f_{n+1}}{f_{n}}\simeq \exp \left[ i\frac{d}{dn}\varphi _n (q_n)
\right].
\end{equation}
Using Eq. (\ref{phi}) and taking into account the fact that
$\partial \varphi _n (q)/\partial q =0$ at $q=q_n$, we obtain:
\begin{equation}\label{tau}
\frac{d}{dn}\varphi _n (q_n)=\omega \int _{q_n}^{a_n}\frac{dq}{v_n(q)}\equiv
\omega \tau _n,
\end{equation}
where $\tau _n$ is the time it takes the oscillator with energy $E_n$
to pass from $q_n$ to $a_n$. Hence, we obtain
\begin{equation}\label{b-fin}
\xi =\cos \omega \tau _n + i\sin \omega \tau _n
=\frac{q_n}{a_n}+i\sqrt{1-\frac{q_n^2}{a_n^2}},
\end{equation}
where
\begin{equation}\label{eq:qnan}
\frac{q_n}{a_n}=\sqrt{\frac{M\omega }{2n+1}}
\frac{|k_n|-\kappa _0}{\kappa _1}.
\end{equation}
We can see that expression (\ref{b-fin}) coincides with $\xi _+$ from
Eq. (\ref{eq:b}) for large $n=N$. This proves that the solution of the
recurrence relation (\ref{eq:f_n}) with the boundary condition
$f_{N+1}^{(+)}/f_{N}^{(+)}=\xi _+$ corresponds to the outgoing wave in the
anion nuclear wavefunction.

If instead of (\ref{eq:chiDA}) we consider the contribution of the
incoming wave in $\chi (q)$, Eq. (\ref{eq:chi}), the coefficients $f_n$
will be given by the complex conjugate of Eqs. (\ref{fn1}), (\ref{eq:fna})
with $A$ replaced by $B$. The corresponding phase factor $f_{n+1}/f_n$
is the complex conjugate of (\ref{b-fin}), asymptotically equal to
$\xi _-=\xi _+^*$ from Eq. (\ref{eq:b}). Hence, the incoming wave contribution
is obtained by solving Eq. (\ref{eq:f_n}) with
$f_{N+1}^{(-)}/f_{N}^{(-)}=\xi _-$.

The relation between the behaviour of $\Psi _{\rm att}({\bf r},q)$ at large
$q$ and the asymptotic form of the amplitudes $f_n$ for closed channels
allows one to find the attachment cross section $\sigma _{\rm att}$
directly from $f_n$. Thus, for the pure outgoing wave case, calculating the
flux for $\chi (q)$ from Eq. (\ref{eq:chiDA}), and using Eq. (\ref{fn1}), we
have:
\begin{equation}\label{eq:DA}
\sigma _{\rm att}=\frac{|A|^2}{k}=2\pi \frac{|f_n|^2\kappa _1}{k \omega }
v(q_n),
\end{equation}
where the last expression is independent of $n$ for large $n$.

Finally, the reflection coefficient $R$ is evaluated by following the
semiclassical wave Eq. (\ref{eq:chiDA}) along the anion potential
curve $U_{0}(q)$ from the matching point $a=q_{\rm m}$ to the right turning
point $b$ and back. Here we assume that due to IVR the potential also acquires
a small negative imaginary part, $-i\gamma /2$. This gives
\begin{equation}\label{R}
R=-i\exp \left[ 2i\int_{q_{\rm m}}^{b}p(q)dq-\gamma \int_{q_{\rm m}}^{b}
\frac{dq}{v(q)} \right],
\end{equation}
where the momentum $p(q)$ and velocity $v(q)$ are calculated as in
Eq. (\ref{eq:chiDA}) using $U_{0}(q)$, and the imaginary part of the
potential results in damping at the rate $\gamma $. If the anion potential
curve for $q>q_{\rm m}$ is described in the harmonic approximation,
\begin{equation}\label{eq:Ua}
U_0(q)=U_a+\frac 12 M\omega _a^2 (q-q_a)^2,
\end{equation}
the integrals in Eq. (\ref{R}) are given by
\begin{eqnarray}\label{eq:pha}
\int_{q_{\rm m}}^{b}p(q)dq &=& \frac{E-U_a}{\omega _a}
\left[ \frac{\pi }{2}-\arcsin \alpha -\alpha \sqrt{1-\alpha ^2}\right],\\
\gamma \int_{q_{\rm m}}^{b}\frac{dq}{v(q)} &=& \frac{\gamma }{\omega _a}
\left[ \frac{\pi }{2}-\arcsin \alpha \right],\label{eq:damp}
\end{eqnarray}
where $\alpha =(q_{\rm m}-q_a)/(b-q_a)$, $b-q_a=\sqrt{(E-U_a)/(M\omega _a^2)}$,
$\omega _a$ is the frequency of the SF$_6^-$ breathing mode, and $|U_a|$ is
the adiabatic electron affinity (neglecting the zero-point energy).

The reflection coefficient that relates the amplitudes corresponding to the
outgoing and incoming waves, $R_f=f_N^{(-)}/f_N^{(+)}$, is obtained with the
help of Eq. (\ref{fn1}), and contains an additional phase factor:
\begin{equation}\label{eq:R_f}
R_f=R\exp \left[ -2i\int _{q_N}^{a_n}p_n(q)dq \right] .
\end{equation}
In the matching procedure the coordinate $q_{\rm m}$ is chosen equal to
$q_N$, Eq. (\ref{eq:q_n}), where $N$ is the truncation number of the
recurrence relation (\ref{eq:f_n}).

\section{Vibrational spectrum density}\label{app:dens}

Let us calculate $\rho (E)$ as the density of states for an ensemble of $s$
harmonic oscillators with frequencies $\omega _i$ for a fixed temperature $T$, 
\begin{equation}
\rho (E)=\frac{e^{S}}{\sqrt{2\pi \langle \Delta E ^2\rangle }},
\end{equation}
where 
\begin{equation}
S=\frac{E}{T}-\sum_{i=1}^s\ln (1-e^{-\omega _i/T})
\end{equation}
is the entropy of the ensemble, 
\begin{equation}
\langle \Delta E^2\rangle =\sum_{i=1}^s\frac{\omega _i^2e^{-\omega _i/T}}
{(1-e^{-\omega _i/T})^2},
\end{equation}
is the variance of the energy, and $T$ is measured in the units of energy.
To find the density for a given energy, the temperature must be chosen
so that the mean energy of the ensemble (measured from the ground state)
is equal to $E$: 
\begin{equation}
E=\sum_{i=1}^s \frac{\omega _i}{e^{\omega _i/T}-1}.
\end{equation}
The density of states calculated in this way is in excellent agreement
with that obtained by directly counting the multimode vibrationally excited
states.



\begin{thebibliography}{99}
\bibitem{comment1} Regarding the behaviour of the calculated SF$_6^-$ potential
curve near the SF$_6$ equilibrium, where it lies above the SF$_6$ curve,
it is worth quoting Ref. \cite{Demkov}: ``There is no justification whatever
for assuming that such a curve determines, for example, the position of
some quasistationary state \dots ''
\bibitem{comment2} Below 90 meV the experimental attachment cross section 
is described accurately by $\sigma _{\rm att}=\sigma /\varepsilon
[1-\exp (-\beta \sqrt{\varepsilon })]$, where
$\sigma =7130~\mbox{\AA }^2\mbox{meV}$,
$\beta =0.405$~meV$^{-1/2}$, and the electron energy $\varepsilon $ is
in meV \cite{KRH92}.
\bibitem{comment3} The anion potential in the harmonic approximation,
Eq. (\ref{eq:Ua}), was described using $U_a=-0.05095$, $\omega _a=0.00325$,
and $q_a=0.275$~a.u., in agreement with the curve from Ref. \cite{Tachikawa}.
\bibitem{comm:ir} Of course, SF$_6^{-*}$ can be stabilised radiatively. The
corresponding rate can be estimated using the dipole transition rates 
of individual infrared-active modes (times their average occupation number
at the energies of SF$_6^{-*}$), and give a lifetime in the tens of
milliseconds, cf. Ref. \cite{Dunning1}.
\end{thebibliography}
\end{document}